\documentclass{ArXiv}

\usepackage{amsmath,natbib}
\RequirePackage{hypernat}
\usepackage[pdftex]{graphicx}
\usepackage{amssymb}
\usepackage{ifthen}
\usepackage{enumitem}
\usepackage{grffile} 
\usepackage{threeparttable} 
\usepackage{url} 

\author{Ulrike Ober}
\address{Georg-August-Universit\"at G\"ottingen,
        Department for Animal Sciences,
        Animal Breeding and Genetics Group, 
        37075 G\"ottingen,
        Germany. E-mail: uober@math.uni-goettingen.de}

\author{Alexander Malinowski}
\address{University of Mannheim, 
  School of Business Informatics and Mathematics,
  A5, 6, 68131 Mannheim,
  Germany.}

\author{Martin Schlather}
\address{University of Mannheim, 
  School of Business Informatics and Mathematics,
  A5, 6, 68131 Mannheim,
  Germany.}

\author[U. Ober, A. Malinowski, M. Schlather, H. Simianer]{Henner Simianer}
\address{Georg-August-Universit\"at G\"ottingen,
         Department for Animal Sciences,
        Animal Breeding and Genetics Group, 
        37075 G\"ottingen,
        Germany.}

\title[The Expected Linkage Disequilibrium in Finite Populations Revisited]{The Expected Linkage Disequilibrium in Finite Populations Revisited}

\usepackage[colorlinks=true,citecolor=blue, linktocpage=true,plainpages=false]{hyperref}

\begin{document}

\begin{abstract}
The expected level of linkage disequilibrium (LD) in a finite ideal population at equilibrium is of relevance for many applications in population and quantitative genetics. Several recursion formulae have been proposed during the last decades, whose derivations mostly contain heuristic parts and therefore remain mathematically questionable.
We propose a more justifiable approach, including an alternative recursion formula for the expected LD.
Since the exact formula depends on the distribution of allele frequencies in a very complicated manner, we suggest an approximate solution and analyze its validity extensively in a simulation study.
Compared to the widely used formula of Sved, the proposed formula performs better for all parameter constellations considered.
We then analyze the expected LD at equilibrium using the theory on discrete-time Markov chains based on the linear recursion formula, with equilibrium being defined as the steady-state of the chain, which finally leads to a formula for the effective population size $N_e$.
An additional analysis considers the effect of non-exactness of a recursion formula on the steady-state, demonstrating that the resulting error in expected LD can be substantial.
In an application to the HapMap data of two human populations we illustrate the dependency of the $N_e$-estimate on the distribution of minor allele frequencies (MAFs), showing that $\hat N_e$ can vary by up to $30 \%$ when a uniform instead of a skewed distribution of MAFs is taken as a basis to select SNPs for the analyses. 
Our analyses provide new insights into the mathematical complexity of the problem studied. 

\keywords{linkage disequilibrium, effective population size, Markov chain, HapMap, allele frequency spectrum}
\end{abstract}

\section{Introduction}

In genetics research, the decay of linkage disequilibrium (LD) as a function of the distance of the considered loci is an important characteristic of a population. 
One measure of LD between two loci which has widely been used in the literature is $r^2$ (cf.\ \citet{HillWeir1994}), which depends on the frequencies of gametes in the considered population.

Moreover, it is commonly assumed that a finite population of size $N$ with constant recombination rate $c$ over time for a given pair of loci achieves a state of ``equilibrium'' after a certain time. Usually, this state of equilibrium is said to be reached when the expected amount of LD does not change from one generation to the next.

The effective population size $N_e$, which is defined as the size of an ideal population at equilibrium with the same structure of LD as the population under consideration (cf.\ \citet{Hedrick2011}), is an important population parameter when considering how real populations evolved over time. 
In practice, $N_e$ cannot be measured but LD can.
Hence, efforts have been made to link the two quantities by formulae of the form $\mathbb{E}(r^2) \approx f(c, N_e)$, with a function $f$ depending on $c$ and $N_e$. 

\subsection{Sved's formula for the expected linkage disequilibrium \citep{Sved1971}}

The following formula for the expected LD in a population was proposed by \citet{Sved1971} and has been used extensively to estimate $N_e$:
\begin{align} \label{SvedUR}
\mathbb{E}(r^2) = \frac{1}{1+4N_ec} \qquad \text{(Sved's formula)}
\end{align}
This equality can be written as
\begin{align*}
 N_e = \frac{\frac{1}{\mathbb{E}(r^2)}-1}{4c} = \frac{1-\mathbb{E}(r^2)}{4c\mathbb{E}(r^2)},
\end{align*}
and by using an empirically estimated $\mathbb{E}(r^2)$, the effective population size $N_e$ can be calculated.
\citet{Hayes2003} argue that the estimated $N_e$ then corresponds to an effective population size $\frac{1}{2c}$ generations ago, if one assumes that the population grows linearly over time.
In the following, we will refer to formula (\ref{SvedUR}) as ``Sved's formula''.
Considering two loci, \citet{Sved1971} derived this formula based on the following recursion formula for the conditional probability $Q_T$ of 
identity by descent (IBD) at the second locus, given that two sampled gametes from the population are IBD at the first locus in generation $T$:
\begin{align}\label{Svedrecursion}
Q_T = \left(1-\frac{1}{2N}\right)(1-c)^2Q_{T-1}+ \frac{1}{2N}(1-c)^2 \qquad \text{(Sved's recursion formula)}
\end{align}
Note that this recursion formula is of linear form $Q_T = aQ_{T-1}+b$ with constants $a$ and $b$.
Sved claims that $Q_T = \mathbb{E}(r_T^2)$, where $r_T^2$ is the LD after $T$ generations.
Additionally, equilibrium is considered to be the point in time for which $Q_{T+1} = Q_T$.
Based on this definition, the equation $Q_T = \mathbb{E}(r_T^2)$ combined with eqn (\ref{Svedrecursion}) yields approximately eqn (\ref{SvedUR}) for small values of $c$ and after replacing $N$ with $N_e$.

Sved's formula has been used in different areas of research and applications, ranging from animal breeding \citep{Meuwissen2001, deRoos2008, Flury2010, Qanbari2010} and plant breeding \citep{Remington2001} to human genetics \citep{Tenesa2007, McEvoy2011}, and it has become one of the standard approaches for $N_e$-estimation.

\subsection{Mathematical shortcomings of previous derivations}

Several other derivations of the formula have been suggested in the last forty years \citep{Sved1973, Tenesa2007, Sved2008, Sved2009}.
We found that all derivations are in some parts of heuristic nature, including mathematical gaps or unsound conclusions.
Indeed, concerns over the validity of the formula and their derivations have already been raised by Sved (cf.\ \citet{Sved2008}, p.\ 185, and a manuscript published on Sved's personal homepage \url{http://www.handsongenetics.com/PIFFLE/LinkageDisequilibrium.pdf}). 
In the following, we will sketch some of the mathematical concerns unfolding in these derivations.

In the manuscript mentioned above Sved reports a misunderstanding in the original derivation \citep{Sved1971}, in which eqn (\ref{Svedrecursion}) is derived, stating that the recursion formula (\ref{Svedrecursion}) should have been
$Q_T = \left(1-\frac{1}{N}\right)(1-c)^2Q_{T-1}+ \frac{1}{N}(1-c)^2.$
But this would not lead to Sved's formula at equilibrium.
It is further argued that a second misunderstanding seems to cancel out the first one leading to eqn (\ref{Svedrecursion}) again, but some uncertainty about the correctness of the equations remains, as stated by Sved in the manuscript mentioned above. 

A second key step in this derivation is the equation $Q_T = \mathbb{E}(r_T^2)$ which finally leads to eqn (\ref{SvedUR}) at equilibrium.
To justify this equation, the following argumentation is used: Imagine, a gamete is sampled at random from the population. A second gamete with the same genotype at the first locus is sampled afterwards. The genes at the first locus are said to be identical by descent (IBD) per definition. 
Then, Sved uses the formula $p_{B_1}^2+p_{B_2}^2$ for the probability of homozygosity at the second locus, where $p_{B_1}$ and $p_{B_2}$ are the corresponding allele frequencies. The expression $p_{B_1}^2+p_{B_2}^2$
is the \emph{unconditional} probability of homozygosity, not taking into account the homozygosity at the first locus in LD,
while the \emph{conditional} probability is expected to be greater than $p_{B_1}^2+p_{B_2}^2$.

\citet{Sved1973} rediscuss this approach
and propose a modified recursion formula which is
$Q_T = \left(1-\frac{1}{2N}\right)(1-c)^2Q_{T-1}+ \frac{1}{2N}$, but the proof of $\mathbb{E}(r_T^2) = Q_T$ is lacking.

Finally, another approach is presented in \citet{Sved2008, Sved2009} by combining the concepts of correlation of two loci and probability of IBD. The critical point in these derivations is that correlations are assumed to be additive. However, this assumption is only verified for the one-locus case, and a proof for the required two-locus case is still missing. 

\citet{Tenesa2007} provide a shorter derivation of Sved's formula using the 
equation $\mathbb{E}(r_{t+1}) = (1-c)r_t$.
Here, the left-hand side is a constant, whereas the right-hand side is a random variable.
Additionally, $\text{Var}(r) \approx \frac{(1-\mathbb{E}(r)^2)}{n}$ is used as a general expression for the sampling variance of an estimate of a correlation coefficient $r$ with sample size $n$. 
In this context, it is not distinguished between the true underlying correlation $\rho$ and the empirical correlation coefficient $r$.
It is not stated either for which underlying distribution this formula can be applied. According to \citet{Hotelling1953}, $\text{Var}(r) \approx \frac{(1-\rho^2)^2}{n}$ holds for a bivariate normal distribution.
Note that the numerator is squared, whereas this is not the case in the formula used by \citet{Tenesa2007}.
It is unclear, whether and how the formula used by \citet{Tenesa2007} is related to the result of \citet{Hotelling1953}, since in the case of LD the underlying distribution is bivariate Bernoulli, and approximation by a bivariate normal distribution is questionable in this case.

All points of critique mentioned so far stress the need for a clearer approach and an extensive empirical analysis of the existing formulae.

\subsection{Organization of the paper}

This paper is organized as follows:
We first propose an alternative linear recursion formula for the expected LD in a finite population and analyze its validity in an extensive simulation study. The new formula is also compared to Sved's recursion formula, and the dependency of the precision of both formulae on the constellation of allele frequencies is analyzed.

We then consider the expected LD at equilibrium in the mathematical framework of the theory on discrete-time Markov chains.
On the basis of a (linear) recursion formula, we derive a formula for the expected amount of LD at equilibrium, leading to a formula for the effective population size $N_e$. First, the derivation is given under the assumption that the recursion formula is exact. We then analyze how the non-exactness of a linear recursion formula affects the result for the expected LD at equilibrium.

In an application, we estimate effective population sizes for the human HapMap data \citep{Hapmap2003} using records of two populations. To illustrate the impact of the allele frequency spectrum used, this is done for different sampling schemes based on minor allele frequencies.

We finally discuss the practical implications of our findings.


\section{Methods and Results}

\subsection{A new recursion formula}

\subsubsection{Basic principles and assumptions:}
\citet{Hill1968} proposed $r^2$ as a measure of LD between a pair of loci. With two biallelic loci $A$ and $B$ with alleles $A_1, A_2, B_1,B_2$ and frequencies $p_{A_1},p_{A_2}, p_{B_1},p_{B_2}$, we denote the frequencies of the genotypes $A_1B_1, A_1B_2, A_2B_1,$ and $A_2B_2$ by $x_{11}, x_{12}, x_{21},$ and $x_{22}$, respectively. Then, 
\begin{align}\label{LD}
r^2 = \frac{(x_{11}x_{22}-x_{12}x_{21})^2}{p_{A_1}p_{A_2}p_{B_1}p_{B_2}}.
\end{align}
Note that if we consider the allelic states at the two loci as Bernoulli variables with parameters $p_{A_1}$ and $p_{B_1}$, then $r^2$ is the square of the correlation coefficient of these two random variables.

In the following, we consider a diploid population of finite size $N$ at some arbitrary point $T=t_0$ in time and two biallelic loci $A$ and $B$ as described above, with gamete frequencies $\mathbf{x}_{t_0} := (x_{t_0,11},x_{t_0,12},x_{t_0,21},x_{t_0,22})$.
Assuming random mating and a constant recombination rate $c$ over time, we can calculate the probabilities $\mathbf{x}_{t_0}^\prime := (x_{t_0,11}^\prime, x_{t_0,12}^\prime, x_{t_0,21}^\prime, x_{t_0,22}^\prime)$ 
for receiving the four different genotypes when producing an offspring gamete as  
\begin{align}\label{newprob}
x_{t_0,11}^\prime = x_{t_0,11} - cD_0, \quad
x_{t_0,12}^\prime &= x_{t_0,12} + cD_0, \quad 
x_{t_0,21}^\prime = x_{t_0,21} + cD_0 \quad \nonumber\\
\text{ and } \quad x_{t_0,22}^\prime &= x_{t_0,22} - cD_0,
\end{align}
with $D_0 := x_{t_0,11}x_{t_0,22}-x_{t_0,12}x_{t_0,21}$. For a detailed derivation we refer to \citet{Hedrick2011}, p.\ $528$ff and the references therein.
We are now interested in the expected squared correlation coefficient $\mathbb{E}_{\mathbf{x}_{t_0}}(r_{t_0+1}^2)$ of the two random variables (the allelic states at the two loci) in $T=t_0+1$, given $\mathbf{x}_{t_0}$ (and hence $r_{t_0}^2$) from $T=t_0$.
If we assume that the population size is constant, the population in $T=t_0+1$ is formed by $2N$ gametes, and the absolute frequencies of the four types of gametes $(n_{11}^\prime, n_{12}^\prime, n_{21}^\prime, n_{22}^\prime) := 2N \mathbf{x}_{t_0+1}$ follow a multinomial distribution with parameters $2N$ and $p = (x_{t_0,11}^\prime, x_{t_0,12}^\prime, x_{t_0,21}^\prime,x_{t_0,22}^\prime)$.

\subsubsection{Analytic expression for the expected LD in the next generation:}

Based on the above assumptions, the exact expected LD in $T=t_0+1$ conditional on $\mathbf{x}_{t_0}$ is given by:  
\begin{align}\label{theo}
\mathbb{E}_{\mathbf{x}_{t_0}}(r_{t_0+1}^2) = \mathbb{E}_{\mathbf{x}_{t_0}}\left(\frac{(n_{11}^\prime n_{22}^\prime-n_{12}^\prime n_{21}^\prime)^2} {(n_{11}^\prime+n_{12}^\prime) (n_{21}^\prime+n_{22}^\prime) (n_{11}^\prime+n_{21}^\prime)(n_{12}^\prime+n_{22}^\prime)}\right),
\end{align}
where $\mathbb{E}_{\mathbf{x}_{t_0}}$ denotes the expectation with respect to the multinomial distribution with parameters $2N$ and $p=\mathbf{x}^\prime_{t_0}$ as described above.
Analytical treatment of this expectation (i.e., expressing it in terms of the probabilities $x_{t_0,ij}^\prime$) does not seem to be feasible for general $N$. The open question is now how to deal with the complex formula.
Even if one tried to approximate the expectation of the ratio by the ratio of expectations (cf.\ e.g.\ \citet{Ohta1971, Hill1977}), the result would still depend on $\mathbf{x}_{t_0}$ in a very complex manner. Therefore, it is reasonable to work with an approximation of this expression, involving only $r_{t_0}^2$ on the right-hand side of eqn (\ref{theo}).

\subsubsection{The alternative recursion formula for the LD:}

According to Sved's approach and based on the assumptions of the previous sections, we propose the following form of an approximate recursion formula for the expected LD in the population, given the gamete frequencies $\mathbf{x}_{t_0}$ in $T=t_0$:
\begin{align}\label{recursion}
\mathbb{E}_{\mathbf{x}_{t_0}}(r_{t_0+1}^2) &= a r_{t_0}^2 + b = ar^2(\mathbf{x}_{t_0}) + b,
\end{align}
where $a$ and $b$ are functions of $c$ and $N$.
Note that $r_{t_0}^2$ is in fact a function of $\mathbf{x}_{t_0}$, which we indicate sometimes by writing $r^2(\mathbf{x}_{t_0})$.
We further choose 
\begin{align}\label{recursion2}
a = (1-c)^2\left(1-\frac{1}{2N}\right) \quad
\text{ and } \quad b &= \frac{1}{2N - 1 - c}.
\end{align}
Note that this choice differs from Sved's recursion formula only in the value of $b$ (cf.\ eqn (\ref{Svedrecursion})) and that we will justify this choice in the subsequent sections.
The coefficients $a$ and $b$ were determined heuristically followed by a systematic validation. 


\subsection{Simulation study to analyze the performance of the new recursion formula}

\subsubsection{Simulation set-up:}

The general idea of the simulation study is the following:
For a given combination $(N, c, \mathbf{x}_{t_0})$ in $T=t_0$, we randomly draw $N_{\text{sample}}$ samples of $2N$ gametes according to the above multinomial distribution with parameters $2N$ and $p = \mathbf{x}_{t_0}^\prime$.
For each of these samples, $\mathbf{x}_{t_0+1}$ and the allele frequencies are obtained as empirical gamete and allele frequencies in $T=t_0+1$, and $r_{t_0+1}^2$ is calculated according to eqn (\ref{LD}). Then, $\mathbb{E}_{\mathbf{x}_{t_0}}(r_{t_0+1}^2)$ is approximated by averaging over the $N_{\text{sample}}$ values of $r_{t_0+1}^2$. Given all tuples $(N,c, r_{t_0}^2, \widehat{\mathbb{E}_{\mathbf{x}_{t_0}}(r_{t_0+1}^2)})$, we can systematically analyze the fit of eqn (\ref{recursion}) in combination with eqn (\ref{recursion2}), as described below.

The simulation was done for all combinations of $N,c$ and $\mathbf{x}_{t_0}$, where
\begin{align*}
N &\in \left\{2^2, 2^3, \ldots, 2^{14} \right\} \allowdisplaybreaks\\
c &\in \left\{0, 0.001, 0.002, \ldots, 0.01, 0.02, \ldots, 0.5\right\}\\
x_{t_0, 11} &\in \left\{0, 0.05, 0.1, \ldots, 1\right\}\allowdisplaybreaks\\
x_{t_0, 12} &\in \left\{0, 0.05, 0.1, \ldots, (1-x_{t_0, 11})\right\}, \text{ for given } x_{t_0, 11}\\
x_{t_0, 21} &\in \left\{0, 0.05, 0.1, \ldots, (1-x_{t_0, 11}-x_{t_0, 12})\right\}, \text{ for given } x_{t_0, 11}, \text{ and } x_{t_0,12}.
\end{align*}
Note that $x_{t_0, 22}$ is determined by $x_{t_0,22} = 1-x_{t_0, 11}-x_{t_0, 12}-x_{t_0, 21}$.
For each parameter constellation, the number of realizations $N_{\text{sample}}$ was chosen dynamically, as described below. 
Further note that $0.05$ was chosen as grid-length for $\mathbf{x}_{t_0}$, although each component of $\mathbf{x}_{t_0}$ can theoretically take only values in $\left\{0, \frac{1}{2N}, \frac{2}{2N}, \ldots, 1\right\}$.
For $N < 32$, we simulated according to this real grid, but got almost identical results.
For large $N$, the computational costs are too high to simulate from the true grid $\left\{0, \frac{1}{2N}, \frac{2}{2N}, \ldots, 1\right\}$.

Parameter constellations causing at least one allele frequency to be zero were excluded from the analyses, 
because in this case $r_{t_0}^2$ cannot be calculated.

\subsubsection{Measures of the goodness of fit:}

We propose the following characteristic as a measure of goodness of fit of eqn (\ref{recursion}):
\begin{align*}
F := \frac{\mathbb{E}_{\mathbf{x}_{t_0}}(r_{t_0+1}^2)-ar_{t_0}^2}{b} - 1, 
\end{align*}
for given $a$ and $b$.
If equalities (\ref{recursion}) and (\ref{recursion2}) were exact, we would observe $F=0$ for all possible parameter constellations.
In the appendix we show that $F$ is closely related to the relative error of the expected LD at equilibrium due to the non-exactness of the recursion formula. 
Note that $F$ is especially sensitive for a misspecification of $b$ in eqn~(\ref{recursion2}).
Since $\mathbb{E}_{\mathbf{x}_{t_0}}(r_{t_0+1}^2)$ is unknown, we use
\begin{align*} 
\hat F = \frac{\widehat{\mathbb{E}_{\mathbf{x}_{t_0}}(r_{t_0+1}^2)}-ar_{t_0}^2}{b} - 1,
\end{align*}
where $\widehat{\mathbb{E}_{\mathbf{x}_{t_0}}(r_{t_0+1}^2)}$ is obtained from the simulation study.

{\textit{Dynamic sampling}:}\\
In the simulation process described above, $N_{\text{sample}}$ was chosen dynamically such that the standard deviation of $\hat F$ was approximately constant over all combinations $(N,c,\mathbf{x}_{t_0})$:
It is 
\begin{align*}
\text{s.d.}(\hat F) &= \text{s.d.}\left(\frac{\widehat{\mathbb{E}_{\mathbf{x}_{t_0}}(r_{t_0+1}^2)}-ar_{t_0}^2}{b}-1\right)\allowdisplaybreaks\\
&= \frac{\text{s.d.}\left(\widehat{\mathbb{E}_{\mathbf{x}_{t_0}}(r_{t_0+1}^2)}\right)}{b}\allowdisplaybreaks\\
&= \frac{\text{s.d.}(r_{t_0+1}^2)}{b \cdot \sqrt{N_{\text{sample}}}}.
\end{align*}
The right-hand side is constant if 
\begin{align*}
N_{\text{sample}} = \left(\frac{\text{s.d.}(r_{t_0+1}^2)}{b}\cdot d\right)^2
\end{align*}
with any constant $d>0$.
To obtain $\text{s.d.}(r_{t_0+1}^2)$, we performed a preliminary simulation study according to the same simulation set-up, but with a constant sample size of $10,000$, and the empirical standard deviation of $r_{t_0+1}^2$ was calculated for all $(N,c,\mathbf{x}_{t_0})$-constellations.
The value of $d$ was chosen such that the maximal value of $N_{\text{sample}}$ equaled $5 \cdot 10^6$.
This led to an average sample size of $462,800$ with a median of $15,000$.

All statistical analyses were performed with R software \citep{RR}.
The R-package ``multicore'' \citep{multicore} was used to parallelize the simulation.

\subsubsection{Results of the simulation study:}

\begin{figure}[!ht]
\begin{center}
\includegraphics[width=\hsize]{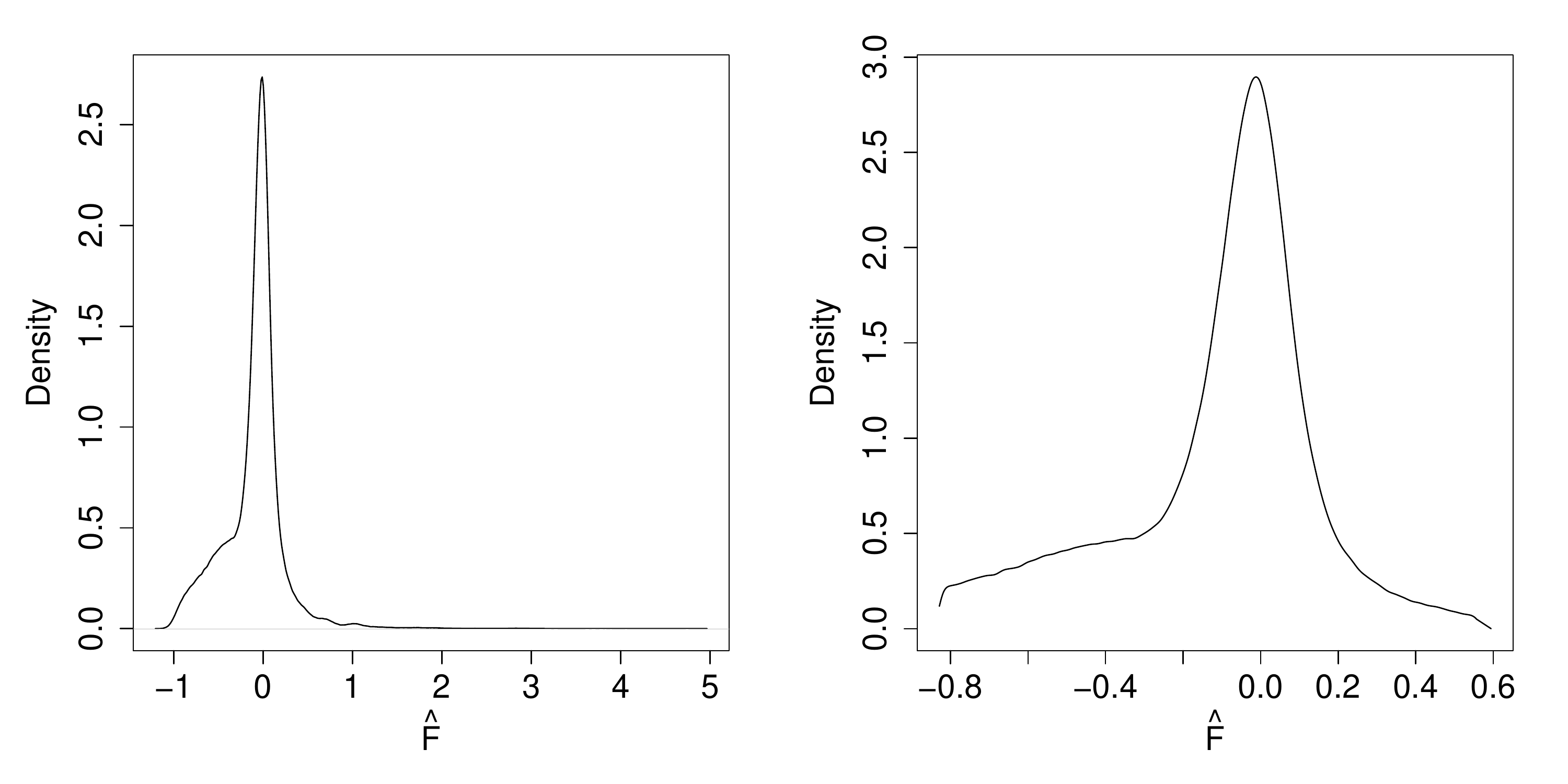}
\end{center}
\caption{Density function of $\hat F$. The left plot shows the density for all obtained values of $\hat F$. To obtain the right density plot, values of $\hat F$ below the $2.5 \%$ quantile and above the $97.5 \%$ quantile of the distribution of $\hat F$ were excluded.}
\label{fig:densityF}
\end{figure}

We found that $\hat F$ was centered around $0$ (Figure \ref{fig:densityF}). When all values of $\hat F$ below the $2.5\%$ and above the $97.5\%$ quantiles were excluded, $\hat F$ ranged between $-1$ and $1$  
indicating that the recursion formula fits the simulated data reasonably well.
Values of $\hat F$ below the $2.5\%$ quantiles and above the $97.5\%$ quantiles were found to be generated by parameter constellations for which 
$$P := x_{t_0, 11}x_{t_0, 12}x_{t_0, 21}x_{t_0, 22}$$
was close to zero, i.e.\ for constellations in which at least one gamete frequency in $T=t_0$ was close to zero (results not shown).

We used boxplots to display $\hat F$ for different parameter constellations.
Boxplots were created separately for different values of $N$, $c$, $P$ and $S :=x_{t_0, 11} + x_{t_0, 12}$. 
Note that $S$ equals the allele frequency $p_{A_1}$ at the first locus and for symmetry arguments is representative for all other allele frequencies.
Values of $P$ (and $S$) were subdivided into $20$ (and $15$) equidistant bins, respectively. 
Outliers (i.e.\ values which lie beyond the extremes of the whiskers) are not displayed in any of the plots.

\begin{figure}[!ht]
\begin{center}
\includegraphics[width=\hsize]{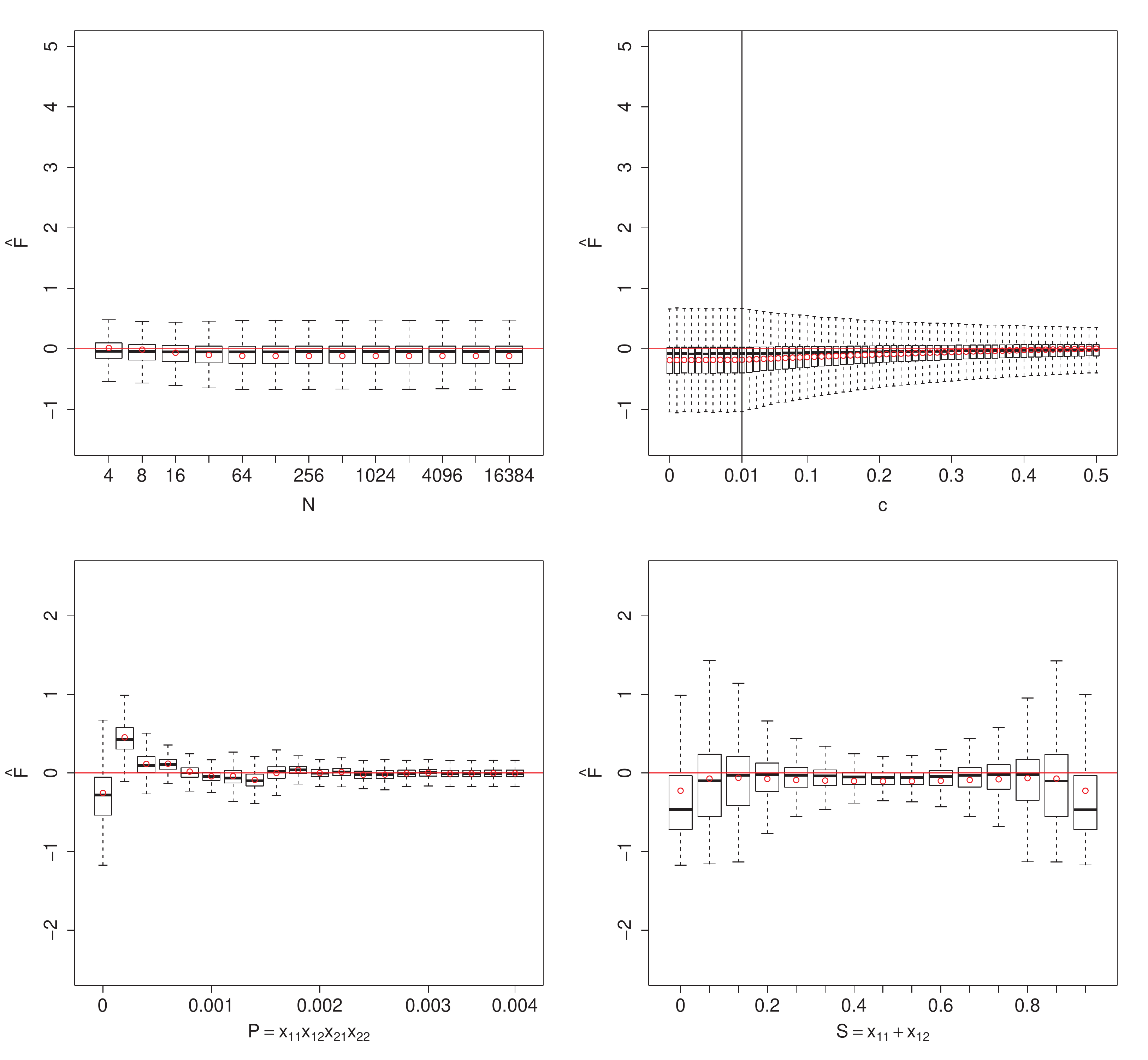}
\end{center}
\caption{Boxplots of $\hat F$, separately for different bins of $N, c, P:=x_{11}x_{12}x_{21}x_{22}$ and $S:=x_{11}+x_{12}$, based on the new recursion formula.
Here, $P$ is the product of gamete frequencies and $S$ is the allele frequency of the first allele at the first locus.
$\hat F$ was calculated according to the new recursion formula. Outliers (i.e.\ values which lie beyond the extremes of the whiskers) are not shown.}
\label{fig:boxplotsNew}
\end{figure}

From Figure \ref{fig:boxplotsNew} we can see that the proposed recursion formula fits the data reasonably well, both for varying $N$ and $c$. The bias as a function of $N$ is almost constant, and it decays with increasing $c$.
The goodness of fit depends heavily on $P$ and $S$: $\hat F$ is larger and more variable for small $P$ and for extreme $S$, but $\hat F$ still ranges between $-1.5$ and $1.5$ in all considered boxplots. 

\begin{figure}[!ht]
\begin{center}
\includegraphics[width=\hsize]{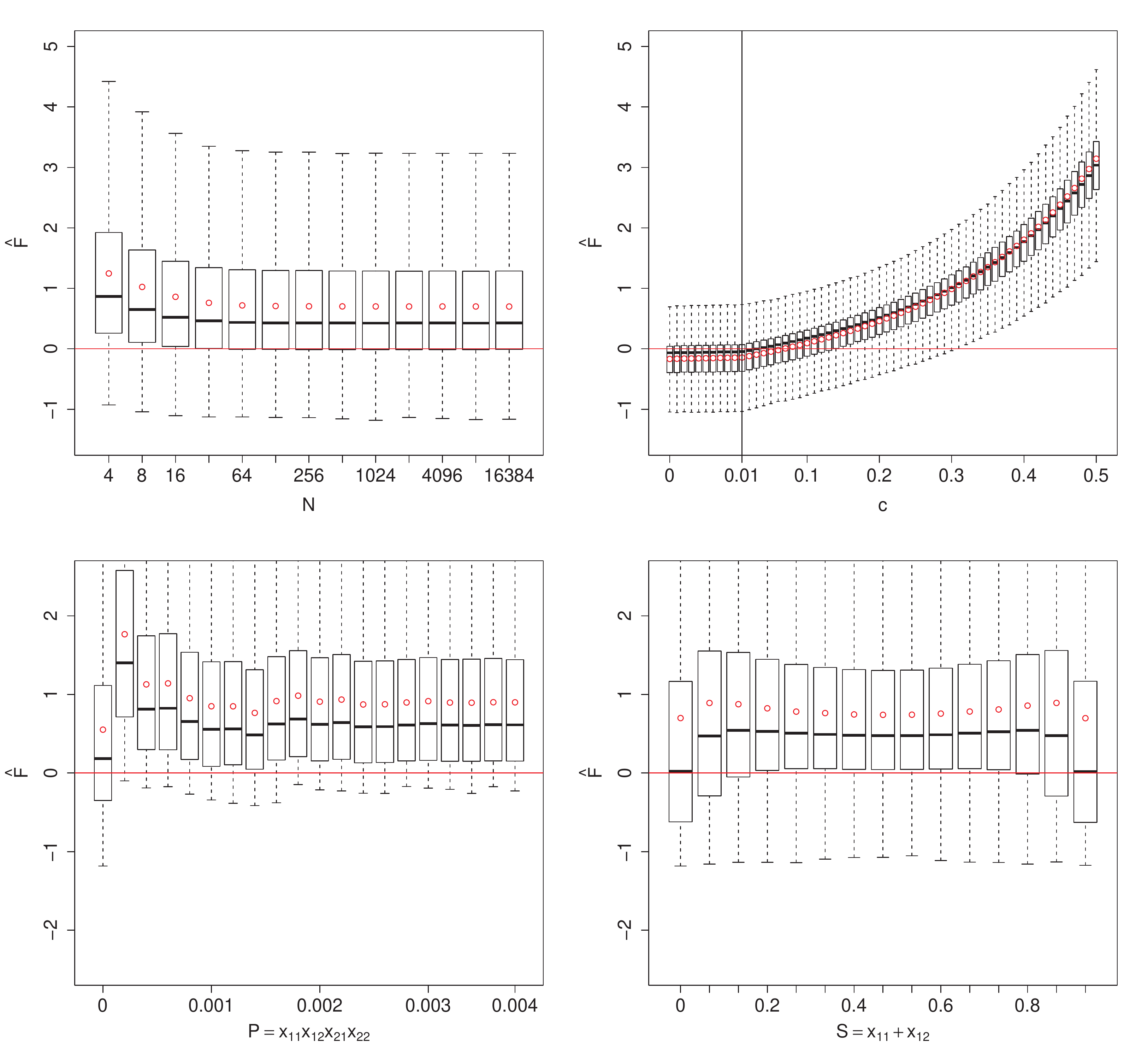}
\end{center}
\caption{Boxplots of $\hat F$, separately for different bins of $N, c, P:=x_{11}x_{12}x_{21}x_{22}$ and $S:=x_{11}+x_{12}$, based on Sved's recursion formula.
Here, $P$ is the product of gamete frequencies and $S$ is the allele frequency of the first allele at the first locus.
$\hat F$ was calculated according to Sved's recursion formula. Outliers (i.e.\ values which lie beyond the extremes of the whiskers) are not shown.} 
\label{fig:boxplotsSved}
\end{figure}

Figure \ref{fig:boxplotsSved} shows that Sved's recursion formula does not fit the simulated data as well as the new formula, especially for $c > 0.01$ and for $N < 30$. This insufficient fit for $c > 0.01$ also pertains to the boxplots for varying $P$ and $S$ (as $\hat F$ is averaged over all constellations of $(N,c,\mathbf{x}_{t_0})$ for a fixed bin of $P$ and $S$, respectively).
For $c < 0.01$, there are only marginal differences between $\hat F$ based on Sved's recursion formula and the new recursion formula.

\begin{figure}[!ht]
\begin{center}
\includegraphics[width=0.5\hsize]{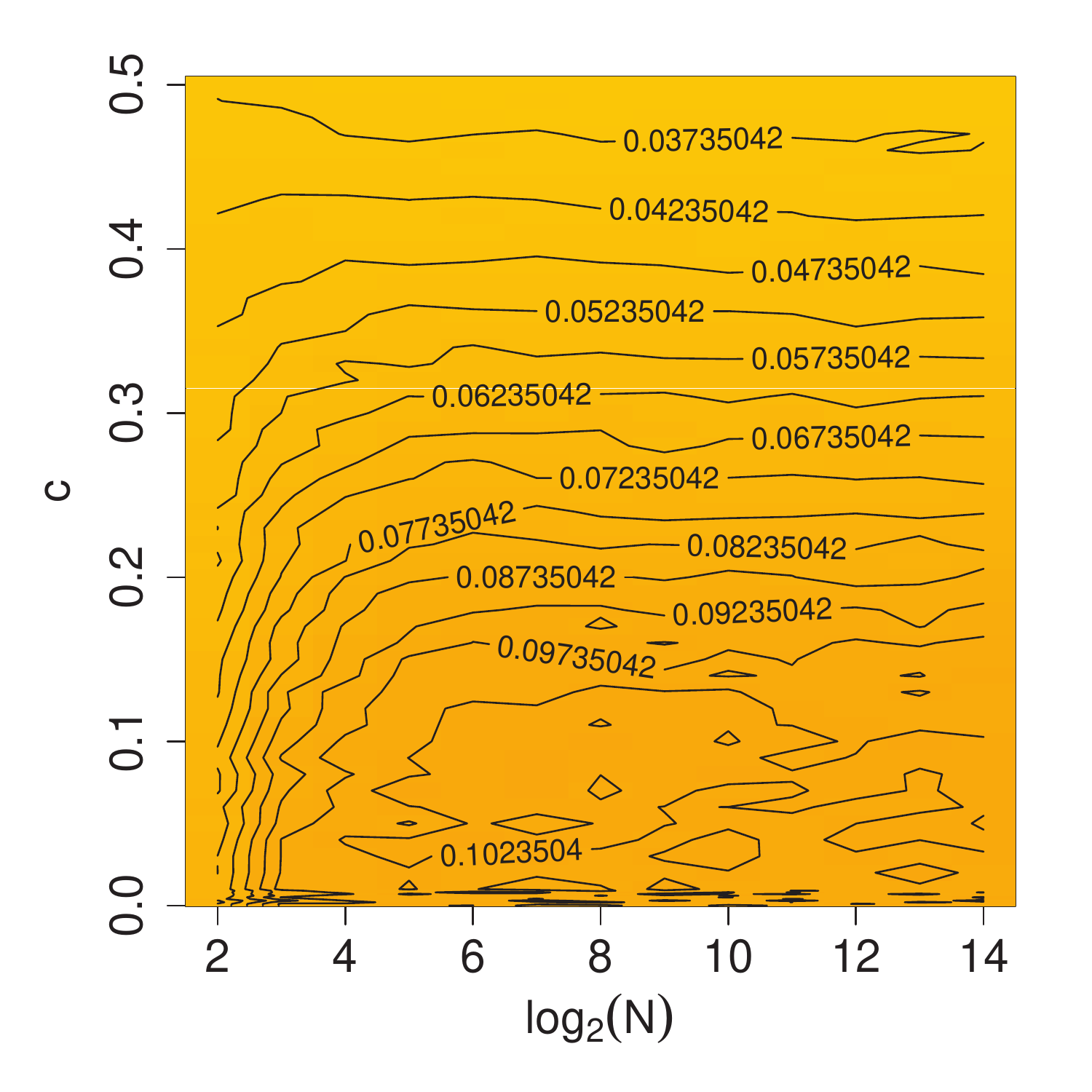}
\end{center}
\caption{Contourplot of the average values of $\hat F^2$.
For a given combination of $(\text{log}_2(N), c)$, the values of $\hat F^2$ were averaged over all possible combinations of $\mathbf{x}_{t_0}$.
Contourplots were created after excluding the extreme $2.5 \%$ quantiles of $\hat F$.}
\label{fig:contourNc}
\end{figure}

Contourplots were drawn for the empirical mean of $\hat F^2$, with the mean calculated using all values of $\hat F$ obtained for a given combination of values on the vertical and horizontal axis of the contourplot.
For example, in the contourplot with axes $(N,c)$ (cf.\ Figure \ref{fig:contourNc}) $\hat F^2$ values were averaged over all possible combinations of $(x_{t_0, 11},x_{t_0, 12},x_{t_0, 21},x_{t_0, 22})$ for a fixed combination $(N,c)$. 
For a clearer representation of the contourplots, we excluded all values of $\hat F$ below the $2.5\%$ and above the $97.5\%$ quantiles beforehand.

\begin{figure}[!ht]
\begin{center}
\includegraphics[width=\hsize]{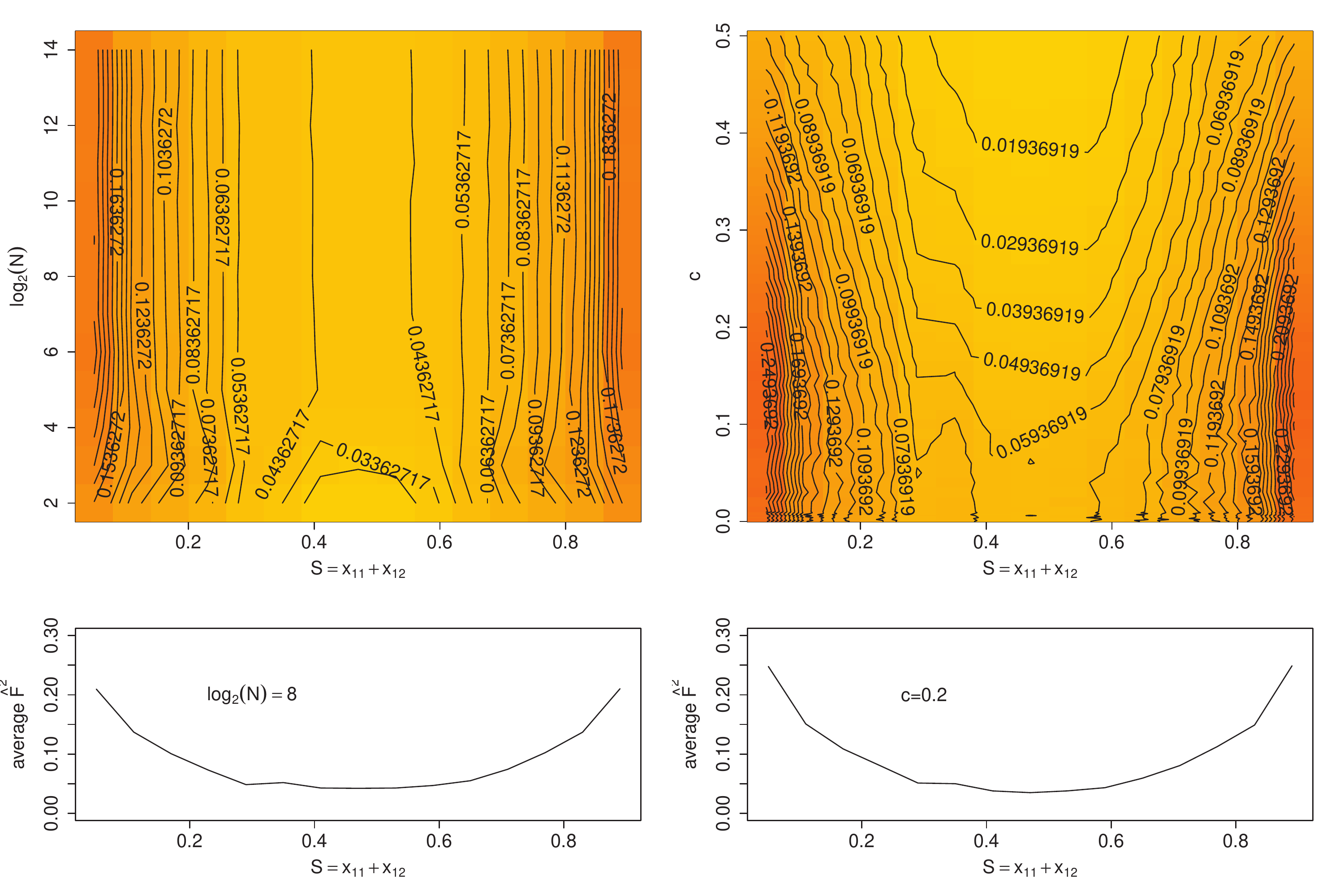}
\end{center}
\caption{Contourplot of the average values of $\hat F^2$.
Left plots: For a given combination of $(S, \text{log}_2(N))$, the values of $\hat F^2$ were averaged over all possible values of $c$ and $\mathbf{x}_{t_0}$  (upper plot).
The lower plot illustrates the average value of $\hat F^2$ as a function of $P$ for $\text{log}_2(N)=8$.
Right plots:
For a given combination of $(S, c)$, the values of $\hat F^2$ were averaged over all possible values of $\text{log}_2(N)$ and $\mathbf{x}_{t_0}$ (upper plot).
The lower plot illustrates the average value of $\hat F$ as a function of $P$ for $c=0.2$.
Contourplots were created after excluding the extreme $2.5 \%$ quantiles of $\hat F$.}
\label{fig:contourS}
\end{figure}

The contourplots of Figure \ref{fig:contourNc} show that $\hat F^2$ depends only slightly on $N$ and $c$, emphasizing the adequate fit of the new recursion formula. Figure \ref{fig:contourS} approves the previous results on the dependency of the goodness of fit on gamete and allele frequencies in $T=t_0$:
The quality of the fit is reduced for $S < 0.2$ and $S > 0.8$.
The same can be observed for $P < 0.0004$ as well as for $\Delta\text{MAF} := \left|\text{min}(p_{A_1}, p_{A_2}) - \text{min}(p_{B_1}, p_{B_2})\right|< 0.2$ (results not shown). 
The term $\Delta\text{MAF}$ describes the absolute difference in minor allele frequencies (MAFs) of both loci, with 
$p_{A_2} = 1-p_{A_1} = x_{t_0, 21}+ x_{t_0, 22}$ and $p_{B_2} = 1-p_{B_1} = x_{t_0,12}+ x_{t_0,22}$. 

Note that similar structures in the contourplots with respect to the dependencies on the allele frequencies can be observed for the goodness of fit of Sved's recursion formula (results not shown).

Overall, the results confirm that an exact recursion formula must depend on the gamete frequencies.
However, this would lead to complex formulae, especially for the state of equilibrium, which we will discuss in the appendix.

\subsubsection{Comparison of slope and intercept of the recursion formulae to empirical values:}

\begin{figure}[!ht]
\begin{center}
\includegraphics[width=0.7\hsize]{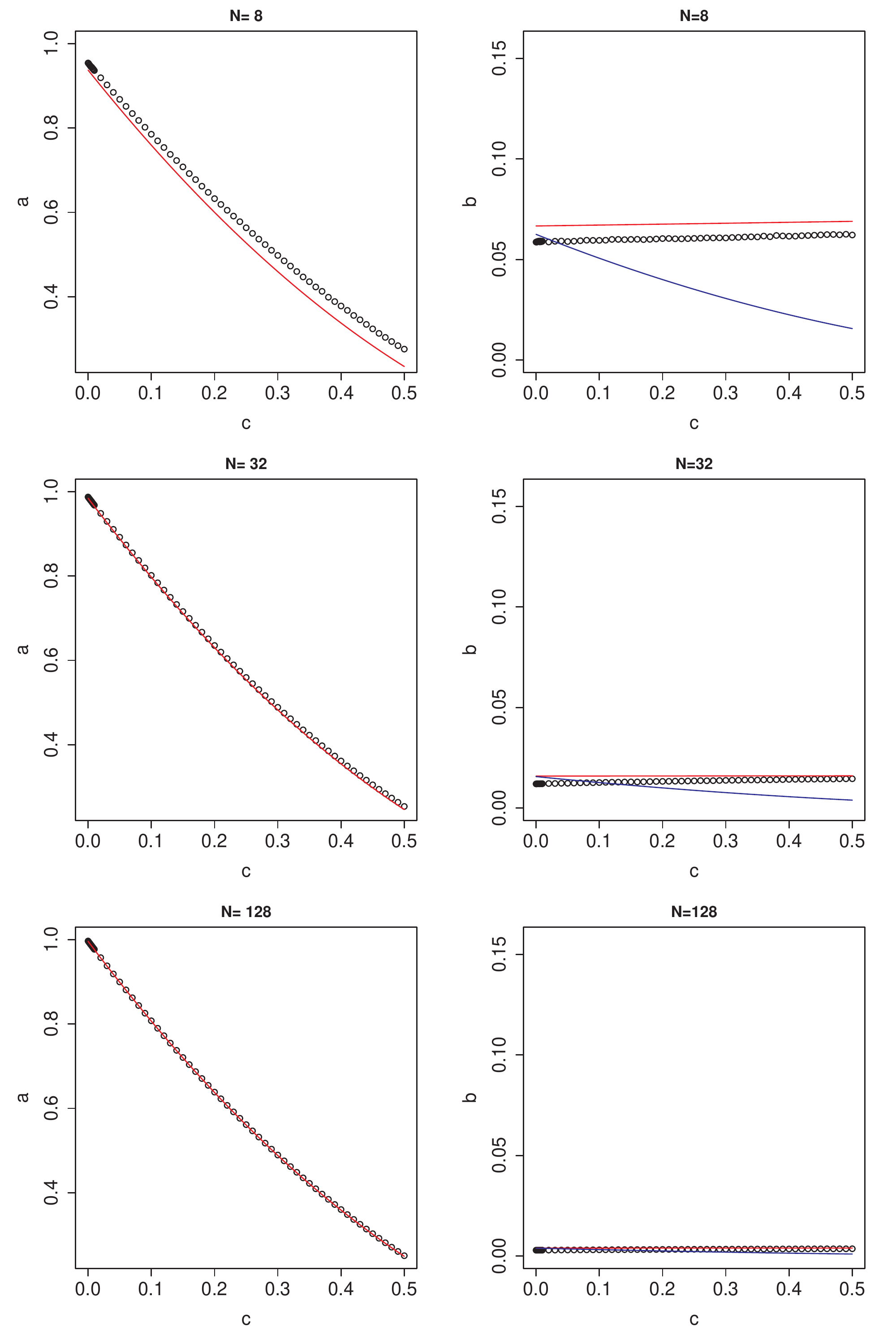}
\end{center}
\caption{Comparison of slope and intercepts of Sved's and the new recursion formula to simulated values.
Left plots: The slope $a$ (identical for Sved's and the new recursion formula) is plotted against $c$ for different values of $N$.
The black dots are the slopes empirically obtained via linear regression.
Hereby, the average LD values obtained in $T=t_0+1$ were regressed against $r_{t_0}^2$.
Right plots: The intercepts $b$ of the recursion formulae are plotted against $c$ for different values of $N$.
Blue (red) lines indicate Sved's (the new) recursion formula.
The black dots are the slopes empirically obtained via linear regression.}
\label{fig:SlopeIntercept}
\end{figure}

We compared the new recursion formula (eqn (\ref{recursion}) in combination with eqn (\ref{recursion2})) to the recursion formula of Sved (eqn (\ref{Svedrecursion})) by plotting the slope $a$ against $c$ for a given $N$. The same was done for the intercept $b$. 
Given $N$ and $c$, we also fitted a linear regression model to the tuples $(r_{t_0}^2, \widehat{\mathbb{E}_{\mathbf{x}_{t_0}}(r_{t_0+1}^2)})$ from the simulation study and added the points $(c, \hat a)$ (and $(c, \hat b)$, respectively) to the plots, where $\hat a$ and $\hat b$ were the estimated slope and intercept from the regression model.
The results are shown in Figure \ref{fig:SlopeIntercept}. 

The slopes $a$ for the two recursion formulae are identical and coincide well with the empirical ones, with a better agreement for larger $N$ (Figure \ref{fig:SlopeIntercept}). The intercepts, though, differ greatly remarkable between the two approaches, especially
for $c > 0.1$ and for small $N$. The intercepts according to Sved's formula are not in agreement with the empirical ones for small $N$ and $c > 0.1$. This is also reflected in the large values of $\hat F$ for increasing $c$ (cf.\ Figure \ref{fig:boxplotsSved}). 
Differences between the intercepts according to Sved's recursion formula and the new recursion formula become less pronounced for increasing $N$ and for decreasing $c$. 

We tried to improve the fit of $a$ and $b$ based on Figure \ref{fig:SlopeIntercept}, especially for small $N$,
by using different formula for $a$ and $b$ in eqn (\ref{recursion2}), e.g.\
$a = \left(1-\frac{1}{3N}\right)\left(1-c\left(1-\frac{1}{3N}\right)\right)^2$ and $b = \frac{1}{2N+1-c}$.
However this did not lead to a significant improvement in terms of $\hat F$ (results not shown).
Since the true relation between $\mathbb{E}(r_{t_0+1}^2)$ and $r_{t_0+1}^2$ is not linear, optimizing a (weighted) average of $\hat F$ is relevant.
Thus, we would recommend to use Figures \ref{fig:boxplotsNew} and \ref{fig:boxplotsSved} as a basis for the assessment of adequacy.


\subsection{The expected LD at equilibrium based on the theory of discrete Markov chains}

\subsubsection{Assuming that the recursion formula is exact:}

In previous studies, ``equilibrium'' was defined as the point in time at which the expected LD of the next generation equals the LD of the previous one (see e.g.\ \citet{Sved1971, Tenesa2007}). Using this definition and assuming a linear recursion formula with coefficients $a$ and $b$ (eqn (\ref{recursion})), the expected LD at equilibrium equals $\frac{b}{1-a}$.

Two major problems arise from this definition:
Firstly, it is not clear whether this equilibrium will ever be achieved.
Secondly, one cannot infer from this definition how the formula for the expected LD at equilibrium is affected if the recursion formula is not exact but only approximate.

To overcome these problems, a mathematically deeper definition of equilibrium can be given based on the theory of Markov chains,
since the sequence of gamete frequencies $\mathbf{x}_T,\, T=t_0, t_0 +1, \ldots$ forms a homogeneous Markov chain with transition probabilities given by the multinomial distribution of the number of gametes of the four types in each generation.
In this framework equilibrium is defined as the steady-state of the considered Markov chain, and the expected LD at equilibrium can be calculated as expectation of $r^2$ under the steady-state distribution.

Under the assumption that the underlying recursion formula is exact, the expected LD at equilibrium based on this approach turns out to be
\begin{align}\label{equi}
R:=\mathbb{E}(r_{\infty}^2) = \frac{b}{1-a}
\end{align}
for $|a| < 1$, in concordance with the above formula.
A detailed derivation based on the Markov chain theory is given in the appendix. 
Note that despite the apparent coincidence with formulae currently used in practice, usually no reference to the Markov chain theory is made. 
The same Markov chain model for the evolution of gamete frequencies has been used by \citet{Karlin1968}, in a study on the ascertainment of fixation probabilities.

Furthermore, the Markov chain theory has the advantage that it also allows the calculation of the expected LD at equilibrium in the case of a non-exact recursion formula. We will come back to this issue in the next section, in which we will analyze how this non-exactness affects the formula for the expected LD at equilibrium.

Using $a$ and $b$ of eqn (\ref{recursion2}) in eqn (\ref{equi}) yields the following formula for the expected LD at equilibrium:
\begin{align}\label{eLD}
R &= \frac{\frac{1}{2N - 1 - c}}{1-(1-c)^2(1-\frac{1}{2N})}\\
  &= \frac{1}{(2N - 1 - c) - (2N - 1 - c)(1-c)^2(1-\frac{1}{2N})} \nonumber
\end{align}
This formula differs from Sved's formula (eqn (\ref{SvedUR})).
Solving eqn (\ref{eLD}) for $N$ yields
\begin{align}\label{Ne}
N = \frac{1}{2(8cR -4c^2R)}\left( Y + \sqrt{-4(8cR-4c^2R)(-R+cR+c^2R-c^3R)+Y^2}\right),
\end{align}
with $Y := 2-2R+8cR-2c^3R$, taking into account that $N$ cannot be negative.

\begin{figure}[!ht]
\begin{center}
\includegraphics[width=\hsize]{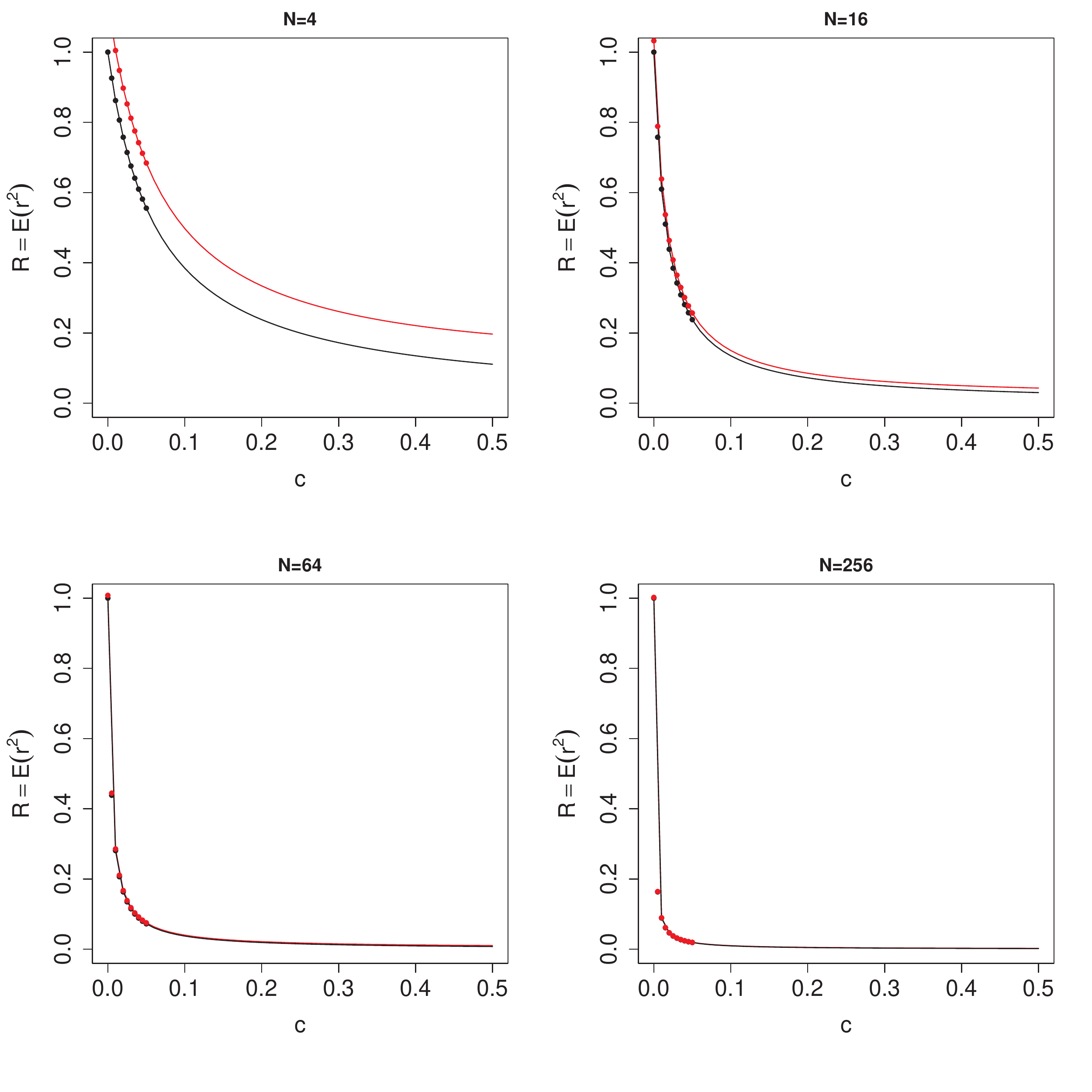}
\end{center}
\caption{Comparison of Sved's formula and the new formula for the expected LD $R$ at equilibrium:
$R$ is plotted against $c$ for different values of $N$.
Red (black) lines show $R$ according to the new (Sved's) formula.
The dots indicate the values of $R$ for $c=0, 0.005, \ldots, 0.05$.}
\label{fig:SvedvsNew}
\end{figure}

To compare the formulae for the expected LD at equilibrium according to eqns (\ref{SvedUR}) and (\ref{eLD}), we plotted $\mathbb{E}(r_{\infty}^2)$ based on both recursion formulae against $c$ for $N \in \left\{4,16,64,256\right\}$. 
Results are shown in Figure \ref{fig:SvedvsNew}. Only small differences between both recursion formulae are observed for large $N$ in Figure \ref{fig:SvedvsNew}, whereas the difference gets more pronounced, when $N$ is small (cf.\ Figure \ref{fig:SlopeIntercept} where similar effects can be realized). The new recursion formula predicts higher values for the expected LD at equilibrium for small values of $N$. 

Real populations usually do not fulfill the implicit assumptions of ideal populations.
It is therefore of interest to calculate the effective population size $N_e$ based on the average LD-value observed from the population for a given value of $c$. By definition, $N_e$ is the size of an ideal population at equilibrium with the same structure of LD as the population under consideration. In practice, $N_e$ is obtained from the right-hand side of eqn (\ref{Ne}), using the average LD-value observed from the data as value of $R$.

\subsubsection{Non-exactness of the recursion formulae:}

As indicated above, another problem which has not been discussed in the literature so far arises from the non-exactness of the recursion formulae.
This problem can also be overcome by the Markov chain theory.

In the case of a non-exact recursion formula, 
the Markov chain theory allows to transfer an error term in the recursion formula to the state of equilibrium.
In the appendix, we provide the corresponding calculations and show that in case of a non-exact recursion formula
\begin{align*}
R^{\boldsymbol{\varepsilon}} := \mathbb{E}_{\boldsymbol{\mu}}(r_{\infty}^2) = \frac{b + \boldsymbol{\pi}^*\boldsymbol{\varepsilon}}{1-a}
\end{align*}
for $|a| < 1$, where $\boldsymbol{\pi}^*$ is the stationary distribution of the considered Markov chain and 
\begin{align*}
\boldsymbol{\varepsilon} &=(\varepsilon(s_1), \ldots, \varepsilon(s_z))\\
\text{with} \quad \varepsilon(s_i) &:= \mathbb{E}_{s_i}(r_{t_0+1}^2) - ar^2(s_i) - b
\end{align*}
is the residual term of the recursion formula depending on the different possible values $s_i$ of $\mathbf{x}_{t_0}$.
Then, the term $\frac{R^{\boldsymbol{\varepsilon}}}{R}-1 = \frac{\boldsymbol{\pi}^*\boldsymbol{\varepsilon}}{b}$ measures the relative influence of $\boldsymbol{\pi}^*\boldsymbol{\varepsilon}$ on the expected LD.

To analyze the effect of the non-exactness, we calculated $\frac{\boldsymbol{\pi}^*\boldsymbol{\varepsilon}}{b}$ for different $(N,c)$-combinations.
For details, we refer to the appendix.
The results are listed in Table \ref{tab:pi_epsilon_durch_b} for $N=4,8,16$ and $c=0.001,0.01,0.05,0.1,0.2,0.3$.
They illustrate that the error-term can lead to a deviance of expected LD up to $25 \%$ suggesting that the effect of non-exactness may be non-negligible.
These analyses were restricted to small values of $N$, since the calculation times increase rapidly with $N$.

\begin{table}[!ht]
\begin{flushleft}
\caption{Values of $\frac{\boldsymbol{\pi}^*\boldsymbol{\varepsilon}}{b}$ for different $(N,c)$-combinations. }
\label{tab:pi_epsilon_durch_b}
\end{flushleft}
\begin{center}
\begin{threeparttable}
\begin{tabular*}{0.6\textwidth}{@{\extracolsep\fill}lrrr}
\hline
$c$ & $N=4$ & $N=8$ & $N=16$\\
\hline
0.001 &-0.264     &-0.199 &-0.067\\
0.01  &-0.255     &-0.165 &-0.023\\
0.05  &-0.176     &-0.088 & 0.096\\
0.1   &-0.116     &-0.018 & 0.195\\
0.2   &-0.074     & 0.053 & 0.297\\
0.3   &-0.025     & 0.097 & 0.324\\
\hline 
\end{tabular*}
\begin{tablenotes}\footnotesize
\item[]Absorbing states were excluded beforehand, and $\boldsymbol{\pi}^*$ was rescaled so that its entries summed up to $1$ afterwards.
\end{tablenotes}
\end{threeparttable}
\end{center}
\end{table}

\begin{figure}[!ht]
\begin{center}
\includegraphics[width=\hsize]{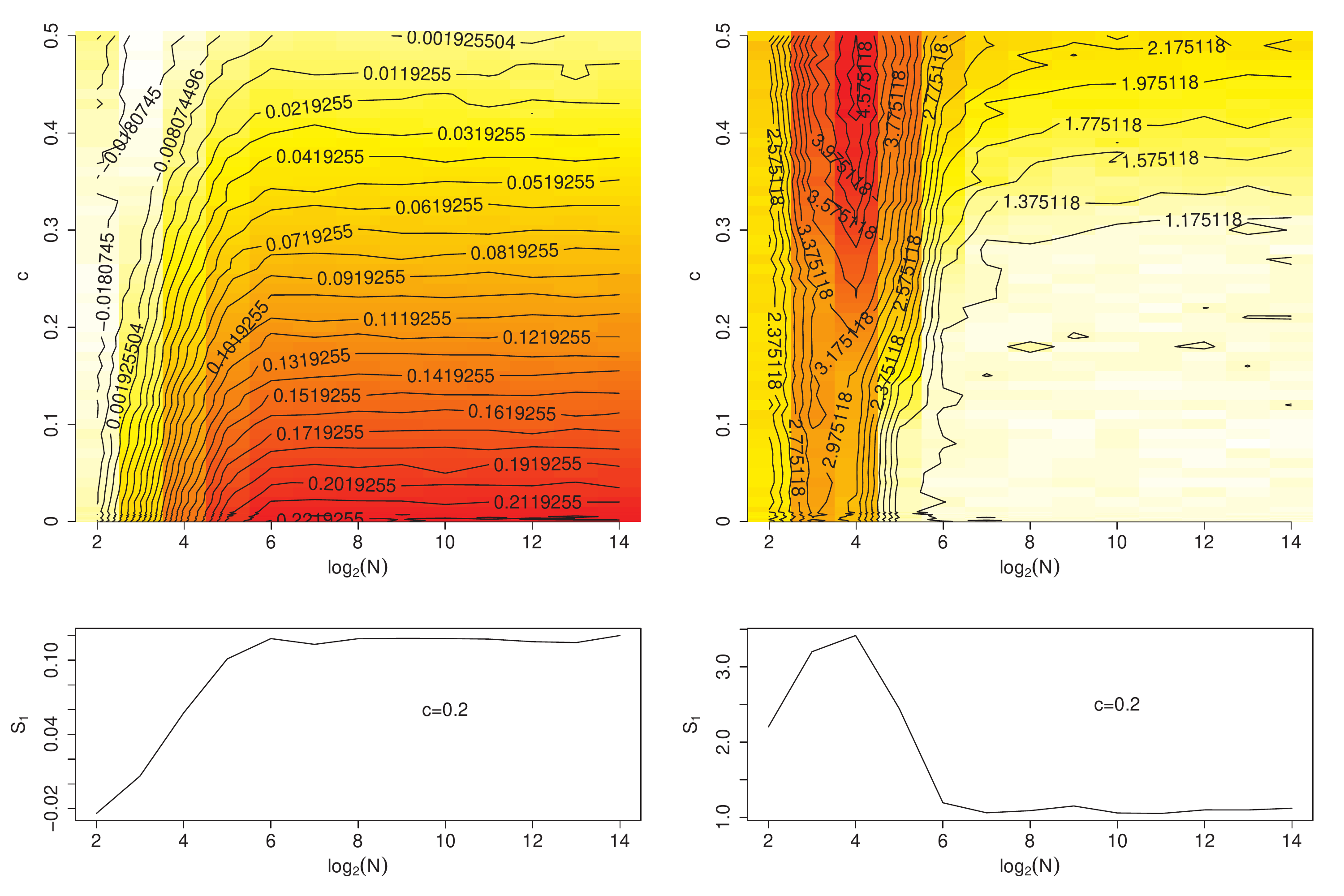}
\end{center}
\caption{Values of $S_1 =  \frac{-\frac{1}{z}\sum \varepsilon_i}{b}$ (upper left plot) and $S_2 = \frac{\max_i |\varepsilon_i|}{b}$ (upper right plot) for different values of $N$ and $c$, calculated on the basis of the simulation study.
 Note that the simulation study was based on a grid of $(x_{t_0,11},x_{t_0,12}, x_{t_0,21},x_{t_0,22})$-values with fixed grid-length $0.05$. Absorbing states were excluded beforehand.
The left (right) lower plot illustrates the values of $S_1 \, (S_2)$ as a function of $\log_2(N)$ for $c=0.2$.}
\label{fig:epislondurchb}
\end{figure}

To get a first impression on the development of $\frac{\boldsymbol{\pi}^*\boldsymbol{\varepsilon}}{b}$ for $N> 16$, we also plotted the negative mean and the maximum value of $\boldsymbol{\varepsilon}$ divided by $b$, given by the terms $S_1 =  \frac{-\frac{1}{z}\sum_i \varepsilon_i}{b}$ and $S_2 = \frac{\max_i |\varepsilon_i|}{b}$, for more values of $N$ and $c$ (Figure \ref{fig:epislondurchb}).
Note that this does not incorporate the stationary distribution $\boldsymbol{\pi}^*$ and that $\boldsymbol{\varepsilon}$-values in these plots were based on the simulation study using the grid of $(x_{11}, x_{12}, x_{21}, x_{22})$-values with fixed grid-distance of $0.05$, which is not as fine as the ``true'' grid if $N > 20$ so that results at this point have to be taken with caution. 
More detailed analyses and a comprehensive simulation study are needed to underpin the quantitative results on the influence of $\boldsymbol{\varepsilon}$ on the expected LD at equilibrium. 

\subsection{Application based on the HapMap data}

As an application of the equilibrium-formula based on the proposed recursion formula, we estimated $N_e$ from LD using human data from the HapMap project \citep{Hapmap2003, Hapmap2007} and applying eqn (\ref{Ne}) as described in the previous sections. We also investigated, how the distribution of MAFs of single nucleotide polymorphisms (SNPs), used to estimate the expected LD in the population, influences the average LD values and hence also the estimates of $N_e$.

\subsubsection{The HapMap data set:}

The HapMap data set comprises $270$ samples from four populations. In this study, we consider two different populations, the Yoruba in Ibadan, Nigeria (YRI) and Utah residents with Northern and Western European ancestry from the CEPH collection (CEU). For each population, the data comprises $30$ trios of individuals.  
The data are available from \url{http://hapmap.ncbi.nlm.nih.gov/downloads/index.html.en}.
For both populations, we used 
allele frequencies from phases II and III (release \#27) 
as well as LD data 
from phases I, II and III (release \#27) for SNPs lying on the $22$ autosomes, and a corresponding genetic map 
(from phase II, estimated from phased haplotypes in release \#22 (NCBI 36)) 
\citep{Hapmap2007}. LD values were available for markers up to $200$kb apart.
For autosome $22$, e.g., there were $\approx 38,000 \, (34,000)$ SNPs occurring in $10,133,060\, (8,130,042)$ LD values for the YRI (CEU) population, for which the genetic distance between the corresponding SNP pairs was available. 
Summing over the $22$ autosomes, there were in total $\approx 2,868,000 \, (2,560,000)$ SNPs occurring in $701,820,000\, (563,239,000)$ LD values for the YRI (CEU) population.

\subsubsection{Estimation of $N_e$ for the YRI and CEU population:}

We estimated $N_e$ separately for the YRI and the CEU population for each of the $22$ autosomes, using eqn (\ref{Ne}) for the expected LD at equilibrium, with $R = \mathbb{E}(r_{\infty}^2)$, estimated as average LD value obtained from the data for given $c$, and replacing $N$ with $N_e$.

Following \citet{Weir1980} we adjusted for the chromosome sample size $n$ by subtracting $\frac{1}{n}$ from the sample-based LD values.
This is necessary, since even in the case of independent loci $\mathbb{E}(r^2)=\frac{1}{n}$.
It has been shown by \citet{Bishop1975}, p.\ 382, that $nr^2$ has an approximate $\chi_1^2$ distribution for a bivariate Bernoulli distribution with independent components, and hence $\mathbb{E}(r^2)=\frac{1}{n}$ in this case.
With this adjustment,
\begin{align}\label{eLDadjusted}
\hat N_e = \frac{1}{2(8c\tilde R -4c^2\tilde R)}\left( \tilde Y + \sqrt{-4(8c\tilde R-4c^2\tilde R)(-\tilde R+c\tilde R+c^2\tilde R-c^3\tilde R)+ \tilde Y^2}\right), 
\end{align}
with $\tilde Y := 2-2\tilde R+8c \tilde R-2c^3 \tilde R$ and $\tilde R = \widehat{\mathbb{E}(r^2)} - \frac{1}{n}$.

\begin{figure}[!ht]
\begin{center}
\includegraphics[width=\hsize]{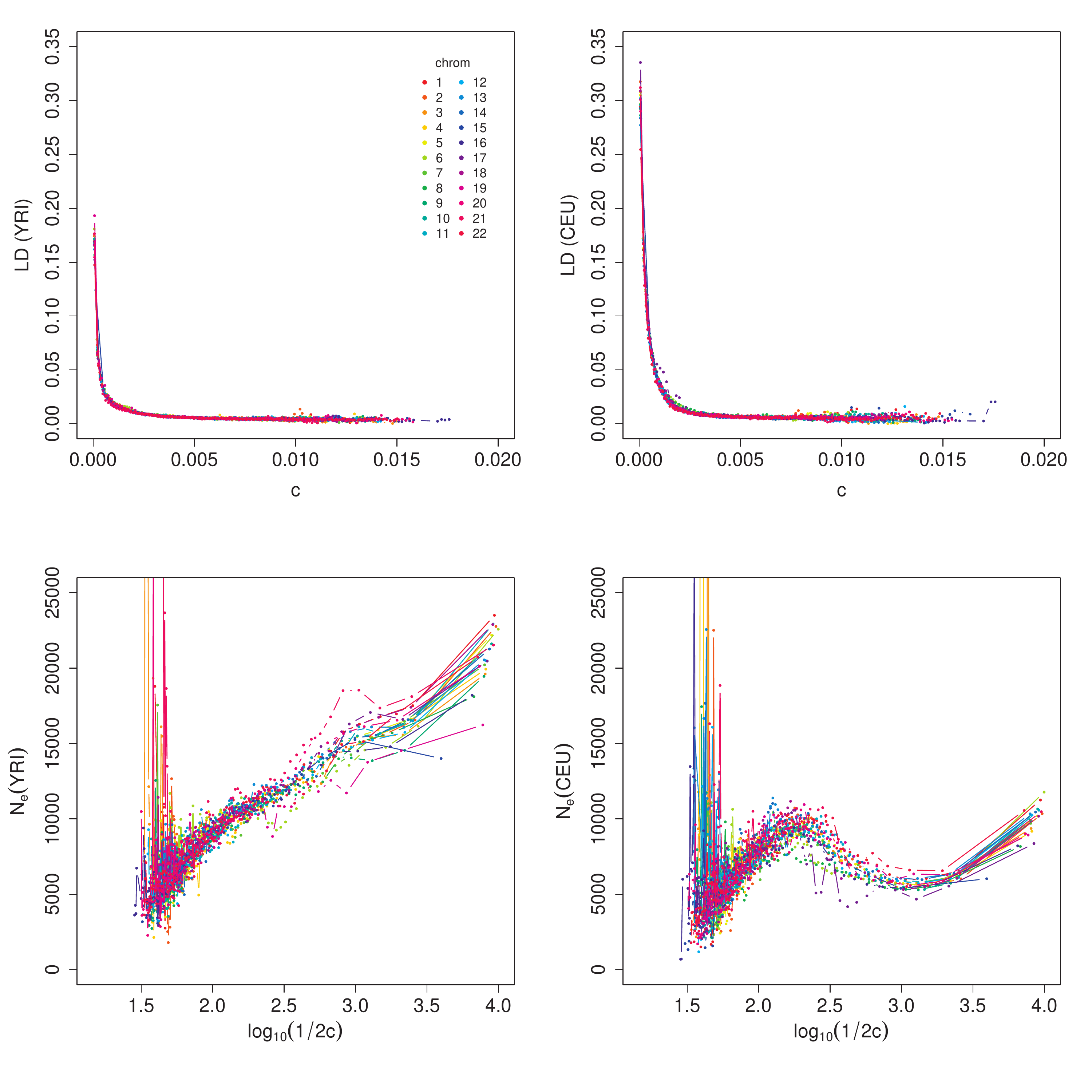}
\end{center}
\caption{LD and estimates of $N_e$ for the YRI and CEU population based on the new recursion formula displayed for the $22$ autosomes.
The upper plots show the decay of LD for varying $c$, estimated from SNPs on different autosomes. 
In the lower plots, the corresponding estimates of $N_e$ are plotted against $\text{log}_{10}(\frac{1}{2c})$.
The left (right) plots are for the YRI (CEU) population. 
$N_e$ estimates based on a given value of $c$ correspond to the point in time ``$\frac{1}{2c}$ generations ago'' \protect\citep{Hayes2003}.}
\label{fig:Hapmap_all_chrom}
\end{figure}

For the HapMap data of the YRI and CEU population, $n=120$, since sequences of $30$ trios for each population were available, comprising $4$ independent parental gametes for each trio.
Estimates of $N_e$ and average LD values were obtained for different bins of the recombination rate $c$.
To classify the pairs of SNPs to the bins, $c$ was approximated by the genetic distance in Morgan. Note that this approach is admissible for small distances. For each autosome, $100$ equidistant bins of $c$ ranging from $0$ to the maximal genetic distance occurring in the data were chosen.
For each bin of $c$, the average $r^2$ value minus $\frac{1}{120}$ was calculated and used in eqn (\ref{eLDadjusted}) to obtain an estimate of $N_e$.
The estimated $N_e$ was plotted against $\text{log}_{10}(\frac{1}{2c})$ (Figure \ref{fig:Hapmap_all_chrom}). 
Additionally, a plot of the adjusted average LD value against $c$ was generated (Figure \ref{fig:Hapmap_all_chrom}).
Results were plotted for all bins of $c$ containing at least $1,000$ LD values.

The decay of LD with genetic distance for the YRI and CEU population can be seen in the upper plots of Figure \ref{fig:Hapmap_all_chrom}, estimates of $N_e$ are displayed in the lower plots.
Note that, since $N_e$ of human populations is large ($N_e > 1,000$), eqns (\ref{eLD}) and (\ref{SvedUR}) basically lead to the same estimates (results not shown).
$N_e$ is smaller for the CEU population (lower right plot of Figure \ref{fig:Hapmap_all_chrom}) and increasing from $\approx 5,000$ to $\approx 10,000$ for $\frac{1}{2c}$ ranging from $1,500$ to $200$, whereas $N_e$ for the YRI population is decreasing for these values of $\frac{1}{2c}$. 
\citet{Hayes2003} argue that the $N_e$ estimate based on Sved's formula (\ref{SvedUR}) for a fixed $c$ corresponds to an estimated $N_e$ approximately $\frac{1}{2c}$ generations ago, if the population grows linearly over time.
Having in mind that the new formula (\ref{eLDadjusted}) and the one based on Sved's formula differ only marginally for large $N$ (cf.\ Figure \ref{fig:SvedvsNew}) and that both the YRI and the CEU populations are large, we will also apply this concept in the following. Assuming a generation interval of $25$ years, the above time frame encompasses $37,500$ to $5,000$ years ago, and we find $\hat N_e \approx 15,000\, (5,800)$ for the YRI (CEU) population $\approx 1,000$ generations ($= 25,000$ years) ago, as well as $\hat N_e \approx 20,500\, (10,000)$ for YRI (CEU) $\approx 8,000$ generations ($= 200,000$ years) ago (cf.\ Figure \ref{fig:Hapmap_all_chrom}). 

For values of $c$ with $\log_{10}(\frac{1}{2c}) < 1.75$, a high variability of $\hat N_e$ values can be observed (Figure \ref{fig:Hapmap_all_chrom}).
We hypothesize that the corresponding values of LD observed from the data are in the order of magnitude one would expect if loci were independent, in which case it would not make sense to estimate $N_e$.
Critical values for $r^2$ could theoretically be derived, if the distribution of $nr^2$ was known for dependent loci (as it is the case for independent loci).

\subsubsection{Influence of distribution of MAFs on the $N_e$-estimates:}

As the detailed analysis of $\hat F$ indicated that the expected LD also depends on the distribution of allele frequencies, it is important to investigate how the underlying distribution of MAFs affects the estimation of $N_e$. 

In commercial SNP array construction for animal breeding purposes, the use of SNPs with uniform MAF distribution is common practice. 
A uniform distribution of MAFs is in general not pursued in human genetics, but may still occur, e.g.\ in studies using phase I data of the HapMap project \citep{Hapmap2003}, where an ascertainment bias can be observed \citep{Nielsen2000, Nielsen2004, Clark2005, Peer2006}.

\begin{figure}[!ht]
\begin{center}
\includegraphics[width=\hsize]{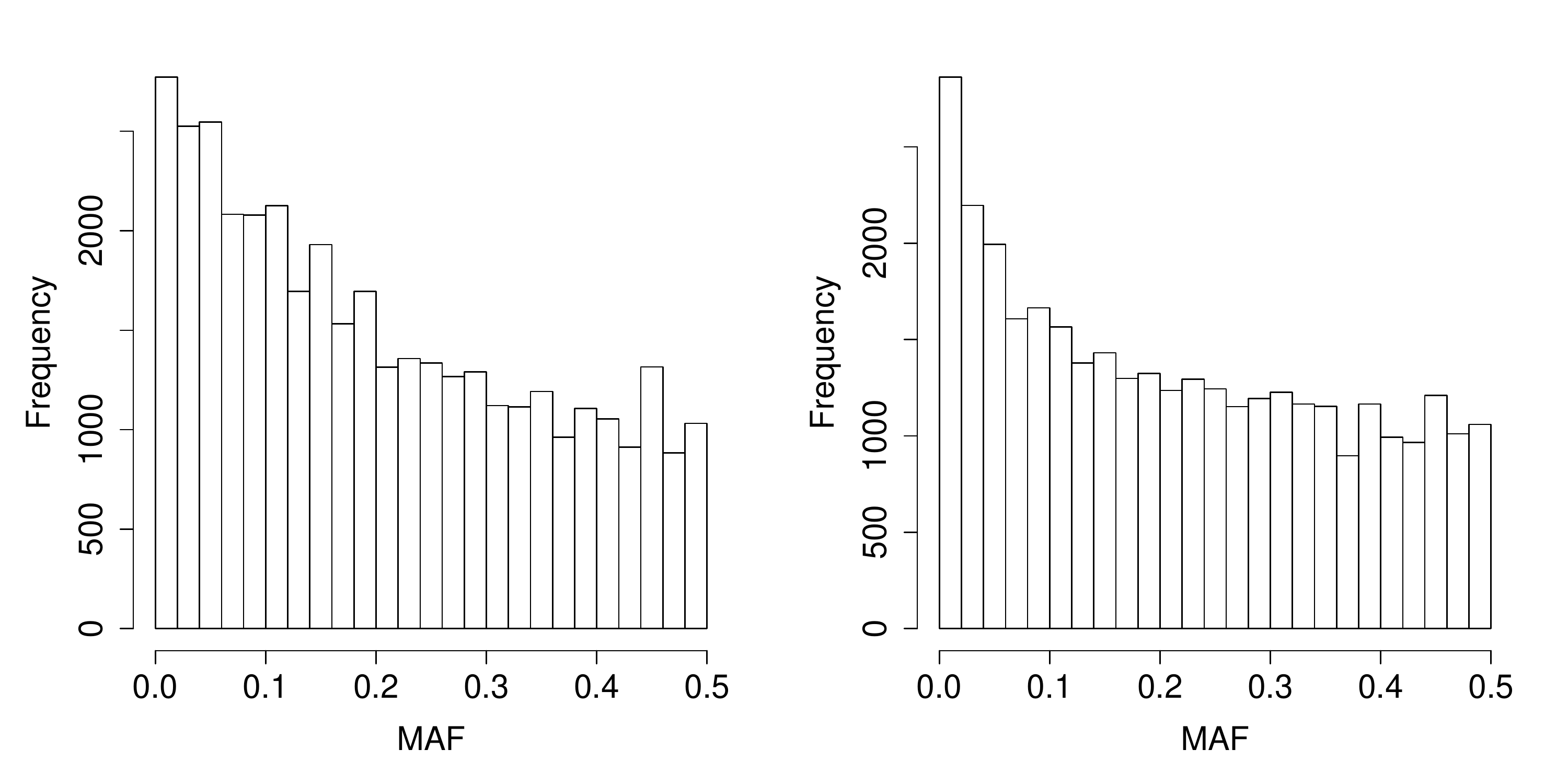}
\end{center}
\caption{Histograms of the distribution of MAFs for all SNPs occurring in the LD data of autosome $22$.
The left (right) plot shows the histogram for the YRI (CEU) population.}
\label{fig:HapmapHistMAF_YRI_and_CEU_single}
\end{figure}

An enforced uniform distribution of MAFs may introduce a systematic and substantial downward bias in $N_e$ estimates, especially for historical effective population sizes, which we demonstrated with the YRI and the CEU population using data of autosome $22$:

Histograms for the MAF values for all SNPs occurring in SNP pairs for which LD values and the genetic distance were available (Figure \ref{fig:HapmapHistMAF_YRI_and_CEU_single}) show that in both populations low MAFs are overrepresented.
For each population we sampled $10,000$ SNP positions out of the $\approx 38,000\, (36,000)$ SNPs available for YRI (CEU) according to two different scenarios: 
\begin{itemize}
 \item[(1)] The $10,000$ positions were sampled randomly, i.e.\ from the true skewed distribution of MAFs
  (cf.\ Figure \ref{fig:HapmapHistMAF_YRI_and_CEU_single}). 
 \item[(2)] MAFs were divided into $10$ equidistant bins between $0$ and $0.5$. Then, $1,000$ SNPs from each bin were sampled to mimic a uniform distribution of MAFs, and all those LD values of pairs of SNPs were kept for which the positions of both SNPs were among the $10,000$ sampled positions.
\end{itemize}

For each scenario, $N_e$ was estimated for different bins of $c$.
We chose $24$ equidistant bins of $c$ ranging from $0$ to $\approx 0.002$ and $25$ equidistant bins of $c$ ranging from $\approx 0.002$ to $\approx 0.02$.
The whole sampling process was repeated $100$ times.
This resulted in $100$ estimated $N_e$ values per scenario and per bin of $c$, for which boxplots were created to display the distributions of $N_e$ for both scenarios.
Boxplots were only created for bins in which on average (averaged with respect to the $100$ replicates) at least $1,000$ LD-values were available.

\begin{figure}[!ht]
\begin{center}
\includegraphics[width=\hsize]{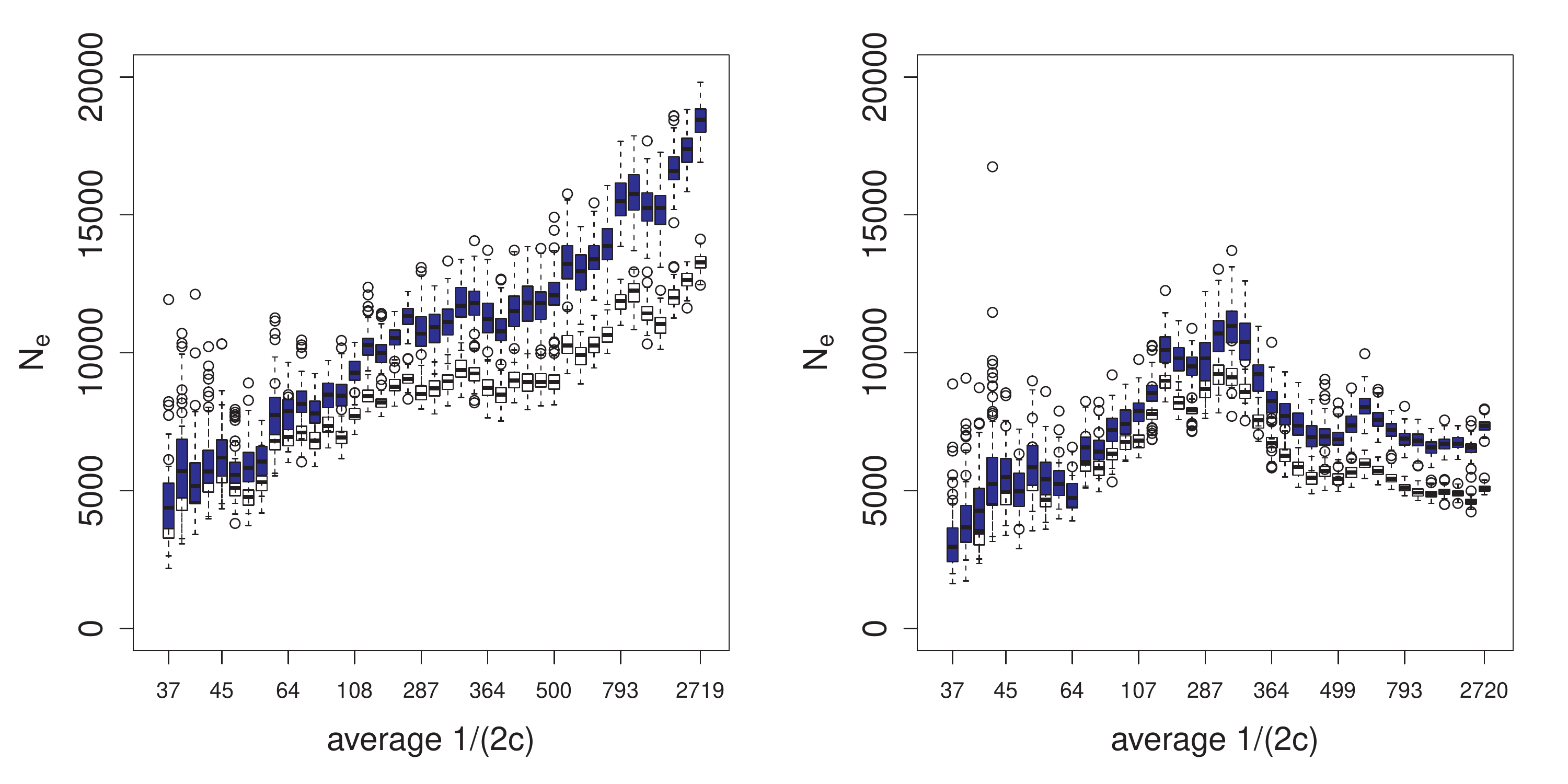}
\end{center}
\caption{Estimates of $N_e$ for the YRI and CEU population for different distributions of MAFs.
The left (right) plot is for the YRI (CEU) population.
Each boxplot represents the distribution of $N_e$ estimates for a given bin of distance $c$ between SNPs in Morgan. 
$N_e$ was estimated based on the new recursion formula.
Estimates of $N_e$ were obtained for two different scenarios: (1) SNP positions were randomly sampled, i.e.\ from a skewed distribution of MAFs.
(2) SNP positions were randomly sampled, so that the distribution of corresponding MAFs was uniform. The sampling process was replicated $100$ times and $N_e$ was estimated for each replicate, resulting in $100$ $N_e$-estimates per scenario and per bin of $c$.
Blue boxplots represent the distribution of $N_e$-estimates for scenario (1), black boxplots represent the distribution of $N_e$-estimates for scenario (2).
Only SNPs on chromosome 22 were considered.
Note that the scale of the x-axis is not linear.
$N_e$ estimates based on a given value of $c$ correspond to the point in time ``$\frac{1}{2c}$ generations ago'' \protect\citep{Hayes2003}.}
\label{fig:HapmapNeMAF_YRI_CEU}
\end{figure}

Figure \ref{fig:HapmapNeMAF_YRI_CEU} illustrates the influence of the distribution of MAFs of SNP pairs used for LD estimation on the $N_e$ estimates in the two populations, respectively.
$N_e$ was estimated for different values of $\frac{1}{2c}$, which can be interpreted as the ``number of generations ago'' in case of Sved's formula \citep{Hayes2003}, as described above.
The plots demonstrate that the $N_e$ estimates using a skewed MAF distribution are up to $30 \%$ larger than the ones using a uniform MAF distribution for large values of $\frac{1}{2c}$.
For example, for $\frac{1}{2c}=500$, the estimated $N_e$ ranges from $\approx 9,000\, (5,400)$ using a skewed MAF distribution to $\approx 12,100\,
(6,900)$ using a uniform MAF distribution and the YRI (CEU) population.


\subsubsection{Comparison of the HapMap estimates with recent results of other studies:}

\citet{Tenesa2007} used the phase I HapMap data to estimate $N_e$ in the YRI and CEU population based on $\approx 1,000,000$ SNPs from $23$ chromosomes. The intermarker distance was in the range of $5$kb to $100$kb for all SNP pairs.
Using only SNPs on autosome $22$ and estimating recombination rates from a nonlinear model, \citet{Tenesa2007} estimated $\hat N_e = 3,246$ for the YRI and $\hat N_e = 1,459$ for the CEU population.
Overall, their estimates appeared to be much lower than the usually quoted value of $10,000$ \citep{Takahata1993, Harding1997}. 
Using a model-free method to estimate recombination rates however changed the estimate of $N_e$ between $+33\%$ and $-45\%$.
Results for the YRI population indicated an ancestral population size of $\approx 7,000$ followed by expansion in the last $20,000$ years ($\approx 1,000$ generations), whereas results for the CEU data supported recent dramatic population growth from an ancestral population size of $\approx 2,500$.

Our study, based on all $22$ autosomes, also indicates a population growth for the CEU population (cf.\ Figure \ref{fig:Hapmap_all_chrom}), whereas no growth can be observed for the YRI population.
One possible reason for the discrepancy of the results may be that \citet{Tenesa2007} used a smaller SNP set and different recombination rates, 
due to the different methods of obtaining these rates.
More importantly, the results were based on a release of phase I, whereas we used phase II data of the HapMap project.
Additionally, \citet{Tenesa2007} excluded all SNPs with MAF $< 0.05$ for the LD estimation, whereas no filtering was performed in the present study. 

\citet{Tenesa2007} also analyzed the effect of a possible ascertainment bias in the HapMap phase I data \citep{Nielsen2000, Nielsen2004, Clark2005, Peer2006} by simulating SNPs with complete ascertainment and simulating SNPs according to a uniform distribution of MAFs (still excluding SNPs with MAF $<0.05$). They found that $N_e$ estimates were biased downwards by $18 \%$ in the second scenario (which is supposed to mimick the MAF distribution of the HapMap phase I data) and concluded that this was also true for their estimated $N_e$ of the HapMap population. 
As opposed to this, in the present study a skewed distribution of MAFs for SNPs occurring in the LD data was observed, as illustrated in Figure \ref{fig:HapmapHistMAF_YRI_and_CEU_single}, which is due to the $\approx 2.1$ million additional SNPs of phase II data compared to the phase I data, which comprised only $1.3$ million SNPs \citep{Hapmap2007}.
However, the simulation results of \citet{Tenesa2007} qualitatively confirm our results of the previous section with respect to the influence of the MAF distribution on the $N_e$-estimates.

\citet{Park2011} used HapMap phase III data to estimate $N_e$ of the current human population based on two different methods, one using the deviation from linkage equilibrium (LE), the other based on the deviation from Hardy-Weinberg Equilibrium (HWE). For the YRI population, estimates fluctuated between $1,275$ and $7,729$, depending on the method, whereas estimates for the CEU population ranged between $1,331$ and $10,437$, illustrating again the great variability of results. \citet{Park2011} argued that the HWE-based method presented the $N_e$ of the current generation, whereas the LE-based method reflected values of ``current and recent'' generations. By considering the ratio of HWE- and LE-based estimates it was found that both populations experienced a recent population growth, which was more distinct for the CEU population.

\citet{McEvoy2011} also estimated $N_e$ based on the HapMap phase III data set using Sved's formula and the concept of \citet{Hayes2003}. It was found that the CEU population experienced a population growth from $N_e \approx 5,000$ to $N_e \approx 11,000$ between $800$ and $240$ generations ago, whereas $N_e$ of the YRI population stayed fairly constant during this time. To decrease a possible ascertainment bias, \citet{McEvoy2011} only used SNPs that were segregating in all populations.

Our estimates of $N_e$ are highly variable in size for recent points in time ($< 50$ generations ago, corresponding to $< 1,250$ years ago). 
A similar variability is reported in \citet{Tenesa2007}, whereas no results are presented in \citet{McEvoy2011} for these points in time.

In summary, our results agree reasonably well with previously reported findings. 
Similar to the findings of \citet{McEvoy2011}, we found the YRI population to be $\approx 2.5$ times as large as the CEU population $1,000$ generations ($25,000$ years) ago, while the effective sizes of the two populations converge when considering more recent points in time. 
The observed increase of the European population between $15,000$ to $10,000$ years before present in the so-called neolithic expansion is in agreement with archaeological findings and coincides well with findings from independent sources, such as the estimations of \citet{Fu2012} based on mitochondrial genomes.
However, it has been shown that the margin of fluctuation is large and that results should always be seen in relation to other existing studies.

\subsubsection{Limitations of the HapMap study:}

As indicated in one of the previous sections, results of the HapMap application have to be taken with caution, since the underlying recursion formula is not exact, contrary to what is assumed in the derivation of formula (\ref{equi}), which underlies formula (\ref{Ne}).
Complications arising from the non-exactness have neither been considered in previous studies so that our results are comparable to the results of other studies from this perspective.
The non-exactness also pertains to the apparent dependency of the development of LD on allele frequencies, for which current recursion formulae do not account.
All findings therefore have to be considered against the background of the implicit assumptions underlying eqn (\ref{Ne}).
Furthermore, we have indicated that the interpretation of different $N_e$-values belonging to different points in the past according to \citet{Hayes2003} had originally been derived for Sved's formula and under the assumption that the population grows linearly over time. In our case, considering sufficiently large populations, formula (\ref{Ne}) differs only marginally from Sved's formula, so that it is reasonable to use the same interpretation, but this approach may not be valid in case of small populations.

%
%

\section{Discussion}


\subsection{The influence of SNP array designs on $N_e$-estimates}

We showed in the simulation study as well as in the application to the HapMap data, that allele frequencies have a strong influence on the performance of the recursion formula and on the estimation of $N_e$.
While the true distribution of MAFs in practical applications (e.g., in sequencing studies) is usually skewed with a substantial excess of small MAF values, commercial SNP arrays are often constructed such that the MAF distribution is uniform, i.e., alleles with extreme MAFs are systematically underrepresented (see e.g.\ \citet{Matukumalli2009} for the construction of a density SNP genotyping array for cattle).
Hence, using LD values based on such an SNP array can have a major impact on estimates of $N_e$ and may result in biased estimates of $N_e$ compared to a situation in which the distribution of MAFs is not uniform.
A similar bias may appear if an SNP array is constructed to reflect the allele frequency spectrum in one population but then is used to estimate $N_e$ in other populations.

\subsection{Analytic expression vs.\ approximate recursion formula}

The proposed recursion formula for $\mathbb{E}_{\mathbf{x}_{t_0}}(r_{t_0+1}^2)$ is still not completely unbiased, which can be seen in the boxplots of Figure \ref{fig:boxplotsNew}.
One possibility to reduce the bias is to use $\tilde{b} = (1+m)b$ instead of $b$ where $m$ is chosen such that 
\begin{align*}
F-m = 0 \quad
\Leftrightarrow \quad \mathbb{E}_{\mathbf{x}_{t_0}}(r_{t_0+1}^2) = ar_{t_0}^2 + \tilde{b}.
\end{align*}
The bias is in fact a function of the gamete frequencies, which can be seen when the upper plots of Figure \ref{fig:boxplotsNew} are created separately for different bins of $P$ and $S$ (results not shown).
This leads back to the problem that an \emph{exact} recursion formula will depend on the frequencies as well.

Even if it was possible to derive an exact recursion for a specific pair of loci with given allele frequencies, many pairs of loci are used to estimate the expected LD, and one would have to account not only for the allele frequencies of a single pair of loci but for the whole distribution of underlying frequencies, which does not seem to be feasible.

\subsection{Obtaining the expected LD at equilibrium directly}

One general way of obtaining the expected LD at equilibrium, without using any recursion formula, is to consider the matrix $\mathbf{P}$ of transition probabilities of the Markov chain and to calculate the limit of $\mathbf{P}^n$ for $n \rightarrow \infty$ to obtain the stationary distribution of gamete frequencies. From this, the expected LD at equilibrium can be calculated directly. 
However, a problem with this approach is that the size of $\mathbf{P}$ is ${2N+3 \choose 2N} \times {2N+3 \choose 2N}$ (there are ${2N+3 \choose 2N}$ possible states of the Markov chain \citep{Karlin1968}),
which makes numerical calculation impossible even for moderately high $N$. We have already encountered this problem when analyzing the term $\frac{\boldsymbol{\pi}^*\boldsymbol{\varepsilon}}{b}$ relating to the non-exactness of the recursion formula.

As an alternative, one could also simulate the Markov chain of gamete frequencies directly (instead of calculating $\mathbf{P}^n$ for $n \rightarrow \infty$) and determine the stationary distribution based on the realizations of the Markov chain. This could be done for different values of $N$ and $c$, and the expected LD could be calculated based on the empirically obtained stationary distribution. Afterwards, the expected LD could be expressed as a function of $N$ and $c$ which then could be used for the estimation of $N_e$. This approach is left for future work.

\subsection{Consequences of non-exact recursion formula}

Previous studies are based on the implicit assumption that the underlying recursion formula is exact and the formula for the expected LD at equilibrium does not incorporate an error-term of the recursion formula.
For $N < 16$, we showed that the error-term in the recursion formula can lead to a non-negligible deviance of expected LD at equilibrium.
These analyses were restricted to small values of $N$ due to the limited calculating capacity and only illustrate the effect qualitatively.
It might be that the error-term becomes negligible for increasing $N$ (and small values of $c$) so that results from previous studies remain reliable.
The critical question remains, how reliable estimates of $N_e$ are if they are based on a non-exact recursion formula, and further research is needed in this field. 

\subsection{Alternative approaches in the literature}

In the literature, there are several other references with alternative approaches to derive formulae for $N_e$ based on LD. 
\citet{Hayes2003}, e.g., state that $N_e$ can be estimated based on the chromosome segment homozygosity (CSH) by using the relation $\text{CSH} = \frac{1}{4N_ec +1}$, which is the same formula one obtains based on Sved's recursion from eqn (\ref{Svedrecursion}) and in their framework, the estimate of $N_e$ corresponds to the point in time ``$\frac{1}{2c}$ generations ago'', as described earlier. However, in the course of their derivation it is assumed that the two considered loci behave independently, which is equivalent to Sved's questionable calculation of homozygosity at the second locus, given the alleles on the first locus are IBD.

\citet{Ohta1971} derived an approximate formula for the expected LD at equilibrium using the theory of diffusion process approximation. Here, the ratio of expectations instead of the expectation of the ratio is used to calculate the expected LD, resulting in
\begin{align}\label{Ohta}
 \mathbb{E}(r^2) \approx \frac{5+2N_ec}{11+26N_ec+8(N_ec)^2}.
\end{align}
\citet{McVean2007} demonstrated that the main difference between Sved's formula for the expected LD at equilibrium and eqn (\ref{Ohta}) is for small values of $N_ec$:
While the expected LD based on Sved's formula approaches $1$ for $c$ tending to zero, eqn (\ref{Ohta}) tends to a value considerably less than $1$.
Comparing both estimates from Monte Carlo coalescent simulation, \citet{McVean2007} found that neither of the formulae provides a particularly accurate prediction for the expected value of LD at equilibrium, unless rare variants (MAF $< 0.1$) are excluded. But eqn (\ref{Ohta}) still predicts the general shape of the decrease in LD with increasing $N_ec$, and it fits the simulated data better than Sved's formula when compared to a sliding average of simulated LD values.

\citet{Song2007} also used diffusion process approximation to derive a formula for the expected LD at equilibrium for a model with recurrent
mutation, genetic drift and recombination. Note that the considered process in diffusion approximation is continuous in both time and space. Diffusion processes possess many nice properties which allow the calculation of certain expectations at stationarity with little effort \citep{Song2007}. \citet{Song2007} were able to express the LD at equilibrium as infinite sum over certain terms, which in turn can be evaluated using the diffusion approximation, finally enabling a numerical calculation of the expected LD. 
One major drawback of this approach is that the diffusion approximation is only valid for sufficiently large populations. For $Nc \rightarrow \infty$ \citet{Song2007} derived a closed-form expression for the expected LD at equilibrium which is the same as obtained by \citet{Ohta1971} for the expectation of the ratio.

\citet{Song2007} provide an approximate formula for the expected LD at equilibrium \emph{directly}, without making a detour via a recursion formula, and despite the fact that derivations based on diffusion approximations are \textit{a priori} valid for sufficiently large populations only, their approach might even constitute a reasonable approximation even for small values of $N$.
So far, we have not compared the validity of the proposed formula for the expected LD at equilibrium in this study with the results obtained by \citet{Song2007}, nor have we compared our $N_e$-estimates with estimates based on coalescent approaches, as \emph{e.g.}\ proposed by \citet{Li2011}, who estimate $N_e$ for all past times via a ``pairwise sequentially Markovian coalescent model''. These comparisons are left for future research.
``Direct'' approaches as applied by \citet{Song2007} or \citet{Li2011} can easily incorporate mutation and recombination rates and 
do not rely on the formula of \citet{Hayes2003} for the determination of the corresponding time in point an $N_e$-estimates refers to, whose derivation was in fact based on the concept of ``chromosome segment homozygosity'' instead of LD.

\subsection{Conclusions}

In this study, we provide a theoretical basis for modeling the evolution of LD in a finite population using the framework of Markov chain theory with underlying multinomial distribution.
On the basis of simulation studies, the HapMap application and the analyses of the state of equilibrium, we can summarize the following important points:

The proposed recursion formula seems to provide a better overall fit than Sved's recursion formula. If $N$ is large or if $c$ is small, differences become marginal. 

The performance of such recursion formulae heavily depends on allele frequencies, and LD is in general a function of the allele frequencies and the gamete frequencies. Hence, estimates of average LD in the population considerably depend on the distribution of MAFs of the SNP pairs used for estimation.
Therefore, if the formula for the expected LD at equilibrium is used to estimate $N_e$, this estimate will also depend on the distribution of MAFs of the SNPs used to calculate the average value of LD for a given genetic distance $c$. This effect was illustrated in the HapMap application.
It is important to keep in mind that
SNP arrays used in certain populations not necessarily will reflect the allele frequency spectrum of this population, which can bias resulting estimates of $N_e$.

The currently used formulae for the expected LD at equilibrium resulting from recursive approaches are based on the assumption that the underlying recursion formulae are correct. As shown in this study, one can theoretically account for the non-exactness of the recursion formula when deriving a formula for the expected LD at equilibrium, but exact solutions in this framework can so far only be obtained for small values of $N$ due to computational limitations. 
For small values of $N$, the expected bias at equilibrium is non-negligible, and we have indicated how the effect can be approximated for larger values of $N$.
Since the effect of the non-exactness might have a substantial influence on the resulting formula, as we have demonstrated in our empirical analyses, this might also be relevant for practical applications. 
In any case, the mathematical complexity of the problem studied warrants some caution when using the results. Estimates of $N_e$ based on this method should always be confirmed by some other, independent method (like proposed by \citet{Song2007}, for instance), and possible sources of bias should critically be monitored.

\section*{Acknowledgments}
This research was funded by the German Federal Ministry of Education and Research (BMBF) within the AgroClustEr `Synbreed -- Synergistic plant and animal breeding' (FKZ 0315528C) in association with the Deutsche Forschungsgemeinschaft (DFG) research training group `Scaling problems in statistics' (RTG 1644).


\bibliographystyle{Chicago}
\renewcommand\refname{Literature Cited} 
\bibliography{diss}

%
%
\section{Appendix}

\subsection{The expected LD at equilibrium based on a recursion formula}

Given a recursion formula like eqn (\ref{recursion}), we will derive a formula for the expected LD at equilibrium which is based on the theory of Markov chains (for introductory books on Markov chains we refer to \citet{Grimmett2001} or \citet{Norris1997}, for instance). Note that the derivation pertains to all recursion formulae with arbitrary coefficients $a$ and $b$ with $|a| < 1$ and that we will provide a general mathematical description of the term ``equilibrium'' which will be defined as the steady-state or ``equilibrium state'' of the considered Markov chain.

According to the multinomial model for the development of the population of gametes, the sequence of gamete frequencies $\mathbf{x}_T,\, T=t_0, t_0 +1, \ldots$ forms a homogeneous Markov chain with transition matrix $\mathbf{P}$ which is given by the multinomial distribution of the number of gametes of the four types in each generation. The parameters of the multinomial distribution are $2N$ and $p = (x_{T,11}^\prime, x_{T,12}^\prime, x_{T,21}^\prime,x_{T,22}^\prime)$.
Since the population size is finite, the Markov chain has a finite set $S$ of states $s_1,\ldots, s_z$.
Here, the $s_i$ are quadruples of frequencies $(x_{11},x_{12},x_{21},x_{22})$, each of which describes a possible partition of the $2N$ gametes into the four types of gametes. In total, there are ${2N+3 \choose 2N}$ possible states \citep{Karlin1968}. 
Let $\boldsymbol{\pi}_T,\, T \geq t_0$, denote the probability vector of $\mathbf{x}_T$. Then
\begin{align*}
\boldsymbol{\pi}_{t_0+n} = \boldsymbol{\pi}_{t_0}\mathbf{P}^n
\end{align*}
for $n = 1, 2, \ldots$.
We write $r_T^2:=r^2(\mathbf{x}_T)$,
$\mathbb{E}_{s_j}(r_{T}^2) := \mathbb{E}(r_{T}^2|\mathbf{x}_{t_0} = s_j)$ for all $T \geq t_0$,
and $\mathbf{e}_j$ for the $j$-th unit vector ($\mathbf{e}_j = (0,\ldots,0,1,0,\ldots,0)$ where the $1$ is at the $j$-th position).

\subsection{Step I: Assuming that the recursion formula is exact}

Let us first assume that the recursion formula (\ref{recursion}) with coefficients $a$ and $b$ holds for some statistic $r^2$ depending on the time $T$ and on the state $\mathbf{x}_{t_0}$ in $T=t_0$.
Note that the following derivation is valid for arbitrary values of $a$ and $b$ with $|a|<1$.
From the recursion formula we get
\begin{align*}
&\sum_j p_j \mathbb{E}_{s_j}(r_{t_0+1}^2) = a\sum_j p_jr^2(s_j)+b
\end{align*}
for all probability vectors $\boldsymbol{\mu}=(p_1, \ldots, p_z)$ with $p_j \geq 0$ and $\sum_j p_j = 1$.
With a slight abuse of notation, we also write $\mathbb{E}_{\boldsymbol{\mu}}(r_T^2)$ for the expectation of $r_T^2$, given that the initial probability vector $\boldsymbol{\pi}_{t_0}$ equals $\boldsymbol{\mu}$.
Then, the last equation is equivalent to 
\begin{align*}
& \sum_j p_j \left(\sum_i (\mathbf{e}_j\mathbf{P})_i r^2(s_i)\right) =  a \mathbb{E}_{\boldsymbol{\mu}}(r_{t_0}^2) + b.
\end{align*}
The left-hand side equals
\begin{align*}
\sum_i \sum_j p_j(\mathbf{e}_j\mathbf{P})_i r^2(s_i) = \sum_i (\boldsymbol{\mu}\mathbf{P})_i r^2(s_i) = \mathbb{E}_{\boldsymbol{\mu}}(r_{t_0+1}^2).
\end{align*}
Hence, we have
\begin{align*}
\mathbb{E}_{\boldsymbol{\mu}}(r_{t_0+1}^2) = a \mathbb{E}_{\boldsymbol{\mu}}(r_{t_0}^2) + b
\end{align*}
for an arbitrary initial probability vector $\boldsymbol{\mu}$, and the weak Markov property yields
\begin{align*}
\mathbb{E}_{\boldsymbol{\mu}}(r_{T+1}^2) = a \mathbb{E}_{\boldsymbol{\mu}}(r_{T}^2) + b
\end{align*}
for all $T \geq t_0$.
If $\boldsymbol{\pi}_{t_0} = \boldsymbol{\mu}$, then $\boldsymbol{\pi}_T = \boldsymbol{\mu}\mathbf{P}^{T-t_0}$,
and the last equation is equivalent to 
\begin{align*}
\sum_j (\boldsymbol{\pi}_{T+1})_j r^2(s_j) = a \sum_j (\boldsymbol{\pi}_{T})_j r^2(s_j) + b.
\end{align*}
If the Markov chain is regular, the convergence theorem for regular discrete Markov chains
yields $\boldsymbol{\pi}_T \rightarrow \boldsymbol{\pi}^*$ for $T \rightarrow \infty$ with $\boldsymbol{\pi}^*$ being the unique stationary distribution,
i.e.\ both sides converge and we get
\begin{align}\label{equi_exact}
\mathbb{E}_{\boldsymbol{\mu}}(r_\infty^2) = a \mathbb{E}_{\boldsymbol{\mu}}(r_\infty^2) + b 
\quad \Leftrightarrow \quad R:=\mathbb{E}_{\boldsymbol{\mu}}(r_{\infty}^2) = \frac{b}{1-a}
\end{align}
for $|a| < 1$.
Note that we need regularity of the Markov chain when applying the convergence theorem and that 
a chain is called ``regular'' if some power of $\mathbf{P}$ contains only positive elements.
In our setting with $\mathbf{P}$ based on the multinomial distribution, this is a priori not true since ``absorbing states'' in the Markov chain exist. These absorbing states reflect situations in which one or more alleles are fixated.
We will deal with this problem later in this appendix.

\subsection{Step II: Dealing with non-exactness of the recursion formula}

Since we know that eqn (\ref{recursion}) with $a$ and $b$ only depending on $N$ and $c$ is not correct, we will now analyze how the non-exactness of eqn (\ref{recursion}) affects the formula $R=\frac{b}{1-a}$ (cf.\ eqn (\ref{equi_exact})).
For each state $\mathbf{x}_{t_0}$, let 
\begin{align}\label{epsilon}
\varepsilon(\mathbf{x}_{t_0}) := \mathbb{E}_{\mathbf{x}_{t_0}}(r_{t_0+1}^2) - ar^2(\mathbf{x}_{t_0}) - b
\end{align}
be the residual term of the proposed recursion formula, and let further
$\boldsymbol{\varepsilon} = (\varepsilon_1, \ldots, \varepsilon_z)^T = (\varepsilon(s_1), \ldots, \varepsilon(s_z))^T$ be the vector of residual terms for the different states $s_i$ of the Markov chain.
As before, we can calculate
\begin{align*}
& \sum_j p_j \left(\sum_i (\mathbf{e}_j\mathbf{P})_i r^2(s_i)\right) =  a \mathbb{E}_{\boldsymbol{\mu}}(r_{t_0}^2) + b + \sum_j p_j\varepsilon(s_j),
\end{align*}
leading to  
\begin{align*}
\mathbb{E}_{\boldsymbol{\mu}}(r_{t_0+1}^2) = a \mathbb{E}_{\boldsymbol{\mu}}(r_{t_0}^2) + b + \boldsymbol{\mu} \boldsymbol{\varepsilon}
\end{align*}
for an arbitrary initial probability vector $\boldsymbol{\mu}$.
The weak Markov property yields
\begin{align*}
\mathbb{E}_{\boldsymbol{\mu}}(r_{T+1}^2) = a \mathbb{E}_{\boldsymbol{\mu}}(r_{T}^2) + b + \boldsymbol{\mu} \mathbf{P}^{T-t_0} \boldsymbol{\varepsilon}
\end{align*}
for all $T \geq t_0$.
If $\boldsymbol{\pi}_{t_0} = \boldsymbol{\mu}$, then $\boldsymbol{\pi}_T = \boldsymbol{\mu}\mathbf{P}^{T-t_0}$,
and the last equation is equivalent to 
\begin{align*}
\sum_j (\boldsymbol{\pi}_{T+1})_j r^2(s_j) = a \sum_j (\boldsymbol{\pi}_{T})_j r^2(s_j) + b + \boldsymbol{\pi}_{T}\boldsymbol{\varepsilon}.
\end{align*}
Using $\boldsymbol{\pi}_T \rightarrow \boldsymbol{\pi}^*$ for $T \rightarrow \infty$ as before, this finally leads
to
\begin{align}\label{equi_nichtexact}
\mathbb{E}_{\boldsymbol{\mu}}(r_\infty^2) = a \mathbb{E}_{\boldsymbol{\mu}}(r_\infty^2) + b + \boldsymbol{\pi}^*\boldsymbol{\varepsilon}
\quad \Leftrightarrow \quad R^{\boldsymbol{\varepsilon}}:=\mathbb{E}_{\boldsymbol{\mu}}(r_{\infty}^2) = \frac{b + \boldsymbol{\pi}^*\boldsymbol{\varepsilon}}{1-a}
\end{align}
for $|a| < 1$.
Hence the formula for the expected LD at equilibrium differs from eqn (\ref{equi_exact}) by the summand $\frac{\boldsymbol{\pi}^*\boldsymbol{\varepsilon}}{1-a}$.

\subsection{Dealing with absorbing states}

In the setting so far, the Markov chain contains several ``absorbing'' states (states which force the chain to move in a certain subset of the set of states).
These absorbing states correspond to situations in which one or two alleles at the considered two loci are already fixed. 
Hence, the Markov chain is not regular and the convergence theorem for Markov chains cannot be applied.
Furthermore, $r^2$ is not defined in case one or more allele frequencies are equal to zero.
In practice, pairs of SNPs with fixed alleles are not considered when estimating the expected LD in the population.
Therefore, we propose to modify the transition matrix $\mathbf{P}$ of the chain by enforcing at least one immediate mutation of an allele in case this allele has become fixed.
The corresponding rows of $\mathbf{P}$ are modified for the absorbing states by choosing the transition probabilities in these rows as indicated in Table \ref{tab:absorbing} mimicking the enforced mutations to leave the absorbing state.

\renewcommand*\thetable{A\arabic{table}}
\setcounter{table}{0}

\begin{table}[!ht]
\begin{flushleft}
\caption{Absorbing states and their modified transition probabilities.}
\label{tab:absorbing}
\end{flushleft}
\begin{center}
\begin{threeparttable}
\begin{tabular*}{\textwidth}{@{\extracolsep\fill}llllcllll}
\hline
\multicolumn{4}{c}{``Absorbing state''} & Transition  & \multicolumn{4}{c}{State after enforced mutation} \\
$x_{11}$ & $x_{12}$& $x_{21}$& $x_{22}$ & probability & $x_{11}$ & $x_{12}$& $x_{21}$& $x_{22}$\\
\hline
1 & 0 & 0 & 0 & 1& $1-\frac{2}{2N}$ & $\frac{1}{2N}$ & $\frac{1}{2N}$ & 0\\
$e^*$ & $1-e$ & 0 & 0 &0.5& $e-\frac{1}{2N}$ & $1-e$ & $\frac{1}{2N}$ & 0\\
    &       &   &   &0.5& $e$ & $1-e-\frac{1}{2N}$ & 0 & $\frac{1}{2N}$\\
\hline
0 & $1$ & 0 & 0 &1&$\frac{1}{2N}$ & $1-\frac{2}{2N}$ & 0 & $\frac{1}{2N}$\\
0 & $e$ & 0 & $1-e$ &0.5&$\frac{1}{2N}$& $e-\frac{1}{2N}$ & 0 & $1-e$\\
  &     &   &       &0.5&0 & $e$ & $\frac{1}{2N}$& $1-e-\frac{1}{2N}$\\
\hline
0 & 0 & 1 & 0  &1&$\frac{1}{2N}$ & 0 & $1-\frac{2}{2N}$ & $\frac{1}{2N}$\\
$e$ & 0 & $1-e$ & 0  &0.5&$e-\frac{1}{2N}$ & $\frac{1}{2N}$ & $1-e$ & 0\\
    &   &       &    &0.5&$e$ & 0 & $1-e-\frac{1}{2N}$ & $\frac{1}{2N}$\\
\hline
0 & 0 & 0 & 1 &1& 0 & $\frac{1}{2N}$ & $\frac{1}{2N}$ & $1-\frac{2}{2N}$\\
0 & 0 & $1-e$ & $e$ &0.5&$\frac{1}{2N}$ & 0 & $e-\frac{1}{2N}$ & $1-e$\\
  &    &      &     &0.5&0 & $\frac{1}{2N}$& $e$ & $1-e-\frac{1}{2N}$\\
\hline 
\end{tabular*}
\begin{tablenotes}\footnotesize
\item[*] with arbitrary constant $e \in \left(0, 1\right)$
\end{tablenotes}
\end{threeparttable}
\end{center}
\end{table}

Note that, since $r_{t_0}^2$ could also not be calculated in the simulation study when one or more allele frequencies were equal to zero, this modification of the Markov chain does not influence the recursion formula and the results of the simulation study with respect to the goodness of fit.

Further note that this modification is biologically inspired by the event of mutations and that this modification is only one possibility among others to deal with the problem of absorbing states.
One could e.g.\ also discard columns and rows of absorbing states in the $\mathbf{P}$-matrix and rescale the rows so that their sums are equal to $1$.
Yet, it is a priori not clear which effect different procedures have on the resulting stationary distribution $\boldsymbol{\pi}^*$ and what they mean in terms of a stochastic model underlying the chain. Further research is needed in this area. 

In the following, we will concentrate on the first possibility described above mimicking biological mutations.
If $c > 0$, all transition probabilities $p_{ij}$ are strictly larger than zero for some power of the modified transition matrix, and the Markov chain is regular allowing for the calculation of expected LD at equilibrium as described in the previous sections.
From now on, we will restrict to the modified transition matrix.

\subsection{The term $\frac{\boldsymbol{\pi}^*\boldsymbol{\varepsilon}}{b}$ as parameter of interest}

Let $R$ be as before and let $R^{\boldsymbol{\varepsilon}}$ denote the expected LD at equilibrium taking into account the error term $\boldsymbol{\varepsilon}$.
Then, the relative difference between these two values is given by
\begin{align*}
\frac{R^{\boldsymbol{\varepsilon}}}{R}-1 = \frac{\frac{b+\boldsymbol{\pi}^* \boldsymbol{\varepsilon}}{1-a}}{{\frac{b}{1-a}}} -1 = \frac{\boldsymbol{\pi}^* \boldsymbol{\varepsilon}}{b}.
\end{align*}
Hence, $\left| \frac{\boldsymbol{\pi}^*\boldsymbol{\varepsilon}}{b} \right|$
measures the relative influence of $\boldsymbol{\pi}^*\boldsymbol{\varepsilon}$ on the expected LD.
Note that $\frac{\boldsymbol{\varepsilon}}{b} = (F(s_1), \ldots, F(s_z))$ and that $\boldsymbol{\pi}^*$ depends on $N$ and $c$.
If we were able to obtain the stationary distribution $\boldsymbol{\pi}^*$ for a fixed combination of $N$ and $c$, we could quantify $\left| \frac{\boldsymbol{\pi}^*\boldsymbol{\varepsilon}}{b}\right|$.
The identity $\frac{\boldsymbol{\varepsilon}}{b} = (F(s_1), \ldots, F(s_z))$ motivates the choice of $F$ as a measure of goodness of fit of the recursion formula since we are especially interested in the expected LD at equilibrium.

The following statistics give a first glance at the behavior of $\frac{\boldsymbol{\pi}^*\boldsymbol{\varepsilon}}{b}$:
\begin{align*}
S_1 :=  \frac{-\frac{1}{z}\sum_i \varepsilon_i}{b} \quad \text{and} \quad S_2 := \frac{\max_i |\varepsilon_i|}{b}
\end{align*}
$S_1$ is closely related to Figure \ref{fig:boxplotsNew} and corresponds to the negative mean of values illustrated in each boxplot.
$S_2$ gives an upper bound for $\left| \frac{\boldsymbol{\pi}^*\boldsymbol{\varepsilon}}{b} \right|$.

\subsection{Empirical analysis based on the new recursion formula}

To analyze the term $\frac{\boldsymbol{\pi}^*\boldsymbol{\varepsilon}}{b}$ for the new recursion formula, we repeated the simulation study described in the Methods section for $N=4,8,16$ and $c=0.001,0.01,0.05,0.1,0.2,0.3$ using the following grid for $\mathbf{x}_{t_0} = (x_{t_0,11},x_{t_0,12},x_{t_0,21},x_{t_0,22})$:
\begin{align*}
x_{t_0, 11} &\in \left\{0, \frac{1}{2N}, \frac{2}{2N}, \ldots, 1\right\}\allowdisplaybreaks\\
x_{t_0, 12} &\in \left\{0, \frac{1}{2N}, \frac{2}{2N},  \ldots, (1-x_{t_0, 11})\right\}, \text{ for given } x_{t_0, 11}\\
x_{t_0, 21} &\in \left\{0, \frac{1}{2N}, \frac{2}{2N},  \ldots, (1-x_{t_0, 11}-x_{t_0, 12})\right\}, \text{ for given } x_{t_0, 11}, \text{ and } x_{t_0,12}.
\end{align*}
As mentioned before, this grid comprises the exact and full set of states of the Markov chain. We chose $N_{\text{sample}}$ so that it had approximately the same magnitude as in the previous simulations.

To obtain the stationary distribution $\boldsymbol{\pi}^*$, we built the transition matrix $\mathbf{P}$ of the Markov chain according to the multinomial distribution. The absorbing states listed in Table \ref{tab:absorbing} were modified as described above.
Then, we calculated $\mathbf{P}^n$ for $n=2^1, \ldots, 2^{15}$.
At equilibrium, each column of $\mathbf{P}^n$ is constant. By graphical inspection, it could be observed that this situation was always reached within $n= 2^{15}$ generations so that all rows of the power $\mathbf{P}^n$ were equal to the stationary distribution $\boldsymbol{\pi}^*$. 
In practice, $\widehat{\mathbb{E}_{\mathbf{x}_{t_0}}(r_{t_0+1}^2)}$ is estimated using SNP pairs with non-fixed alleles in the population.
Hence, we are interested in $\frac{\boldsymbol{\pi}^*\boldsymbol{\varepsilon}}{b}$ where $\boldsymbol{\pi}^*$ and $\boldsymbol{\varepsilon}$ only contain non-absorbing states.
Therefore, we calculated $\varepsilon_i(\mathbf{x}_{t_0})$ for all non-absorbing states $\mathbf{x}_{t_0}$ based on $\widehat{\mathbb{E}_{\mathbf{x}_{t_0}}(r_{t_0+1}^2)}$ obtained from the simulation and rescaled $\boldsymbol{\pi}^*$ so that its sum equaled $1$ after excluding all absorbing states. Then, $\frac{\boldsymbol{\pi}^*\boldsymbol{\varepsilon}}{b}$ could be calculated for different $(N,c)$-combinations, to analyze the influence of the non-exactness of the recursion formula.

\end{document}